\documentclass[groupeaddress,aps,article,nofootinbib,10pt]{revtex4-2}
\usepackage{array}
\usepackage{titlesec}
\titleformat{\paragraph}
{\normalfont\normalsize\bfseries}{\theparagraph}{1em}{}
\titlespacing*{\paragraph}
{0pt}{3.25ex plus 1ex minus .2ex}{1.5ex plus .2ex}
\renewcommand*{\theparagraph}{\roman{paragraph})}

\newcommand{\vb}{\mathfrak{e}}
\newcommand{\vbt}{\mathfrak{e}^{0}_{~t}}

\usepackage{amssymb, amsmath,amsthm,bm}    
\usepackage{graphicx}   
\usepackage{verbatim}   
\usepackage{color}      
\usepackage[labelformat=simple]{subcaption}
\usepackage{mathrsfs} 
\usepackage{dsfont}
\usepackage[linktocpage = true]{hyperref}  
\usepackage{cancel}
\usepackage{graphicx}
\usepackage{caption} 
\usepackage{tcolorbox}
\usepackage{yfonts}
\usepackage{multirow}
\definecolor{green}{RGB}{35,142,35}
\usepackage{float}
\usepackage{upgreek}

\usepackage{empheq} 
 
\usepackage{ulem}

\usepackage[framemethod=tikz]{mdframed}

\hypersetup{
    colorlinks   = true,
    linkcolor = blue,
    citecolor    = red
}  
\makeatletter
\def\p@subsection{}
\makeatother

\usepackage{cleveref}

\allowdisplaybreaks

\label{eq:partition_function_general2}

\newtcolorbox{mymathbox}[1][]{colback=white, #1}

\usepackage[framemethod=tikz]{mdframed}
\renewcommand{\FR}[2]{\frac{#1}{#2}}

\newcommand{\ncmd}{\newcommand}
\ncmd{\lt}{\left}
\ncmd{\rt}{\right}
\newcommand{\eq}[1]{Eq. \eqref{#1}}
\newcommand{\fig}[1]{Fig. \ref{#1}}
\newcommand{\tr}[1]{\mbox{Tr}\lt[{#1}\rt]}

\ncmd{\kF}{$k_F$ }
\ncmd{\Lf}{$\Lambda_f$ }
\ncmd{\Lb}{$\Lambda_b$ }
\ncmd{\KF}{k_{\mathrm{F}}}

\definecolor{darkblue}{RGB}{10,10,150}

\DeclareMathOperator{\Ei}{Ei}

\newcommand{\dd}{\mathrm{d}}

\makeatletter
\newcommand*{\rom}[1]{\expandafter\@slowromancap\romannumeral #1@}
\makeatother

\makeatletter
\renewcommand\theequation{{\color{blue} \theequation@prefix \arabic{equation}}}
\makeatother

\makeatletter
\def\p@subsection{}
\makeatother

\newcommand{\bqa}{\begin{eqnarray}} 
\newcommand{\eqa}{\end{eqnarray}}
\newcommand{\nn}{\nonumber \\}

\newcommand{\spmqty}[1]{\begin{psmallmatrix}#1\end{psmallmatrix}}

\newcommand{\abs}[1]{\left| #1 \right|}

\begin{document}

\title{
Emergence of curved momentum-spacetime and 
its effect on
the cyclotron motion 
\\
in the antiferromagnetic quantum critical metal 
}

        \author{Francisco Borges$^{1,2}$\footnote{fborges@perimeterinstitute.ca} }
        \author{Sung-Sik Lee$^{1,2}$\footnote{slee@mcmaster.ca}
        }
        \affiliation{$^{1}$Department of Physics \& Astronomy, McMaster University, Hamilton ON L8S 4M1, Canada}
        \affiliation{$^{2}$Perimeter Institute for Theoretical Physics, Waterloo ON N2L 2Y5, Canada}
    
        \date{\today}
        
\begin{abstract}
We show that anisotropic quantum corrections
 can dynamically give rise to curved momentum-spacetimes for quasiparticles in metals.
In the (2+1)-dimensional antiferromagnetic quantum critical metal, 
a curved momentum-spacetime arises as
the critical spin fluctuations generate red shift that dilates frequency of electron unevenly on the Fermi surface.
As the disparity of the momentum-dependent red shift is controlled by the shape of the Fermi surface, 
the momentum-spacetime geometry that emerges at low energies depends on the  bare nesting angle of the Fermi surface.
With increasing  nesting angle,
the region in which electron motion is slowed down
by critical spin fluctuations shrinks.
On the other hand, the increasing nesting angle 
makes the red shift stronger near the hot spots due to the weakened
screening of the interaction.
These competing effects result in a non-monotonic dependence of the cyclotron frequency of electron on the nesting angle of the Fermi surface.
The red shift that becomes more singular at the hot spots with increasing nesting angle
creates a possibility of realizing a momentum-space 
black hole horizon beyond a critical nesting angle
 : the electron motion becomes `perpetually' slowed down 
 as it approaches a hot spot
 in the same way that 
 the motion of a free falling object 
 freezes near the event horizon of a black hole 
 with respect to an asymptotic observer.
However, the analogous horizon 
 in momentum space does not lead to a vanishing cyclotron 
frequency because the metric singularity
 at the hot spots 
is cut off by
thermal effects
 present above the non-zero superconducting transition temperature.

\end{abstract}
\maketitle
\onecolumngrid

\section{Introduction}

The semi-classical equation of motion of quasiparticles in solids 
is remarkably symmetric 
under the interchange of position and momentum.
The momentum-dependent quasiparticle energy is the counter part of the position-dependent potential in real space. 
The Berry curvature associated with the Bloch wavefunctions plays the role of the magnetic field in momentum space\cite{Xiao_2010}.
It is then natural to ask if the symmetry can be further extended to spacetime geometry\cite{crnkovic1988symplectic,KOWALSKI_GLIKMAN_2013,doi:10.1063/1.528801,ASHTEKAR1991417,Davis_2022}.
A real-space curvature can be created through
buckling of lattices 
or topological defects in solids\cite{Amorim_2016,WANG2021113204,pereira2010geometry,Mao_2020,Vozmediano_2010,CORTIJO2007293,Wilde_2021,PhysRevLett.130.066001,PhysRevB.104.L201107}. 
Recent studies on the semi-classical equations of motion of quasiparticles have suggested the emergence of a curved momentum space in lattice models.
In particular,
it has been pointed out that the non-linear response of quasiparticles to the external electromagnetic field 
 can be captured 
 geometrically\cite{
 smith2021momentumspace,holder2021electrons}.
%
%
In those examples,
the metric\cite{RevModPhys.84.1419} is ultimately determined from the 
single-particle wavefunction fixed by the underlying lattice.
%
%
%
In this paper, we consider an intrinsic physical mechanism by which momentum space and time is integrated into a curved {\it momentum-spacetime}  through the electron-electron interactions.
We point out that curved momentum-spacetimes 
naturally arise from anisotropic quantum corrections in metals
and 
even a momentum-space black hole horizon can emerge
if quantum corrections are  strongly singular in momentum space.
%

Strongly momentum-dependent quantum corrections arise in metals close to quantum critical points associated with order parameters carrying non-zero momenta.
At spin or charge density-wave critical points, electrons residing on hot manifolds, sub-manifolds of Fermi surface that can be connected by the ordering wave vectors,
are more strongly scattered by critical fluctuations than electrons away from the hot manifolds.
This leads to a momentum-dependent renormalization of electron.
In particular, the Fermi velocity can acquire a strong dependence
 on momentum along the Fermi surface 
 as electrons become significantly heavier 
 near the hot manifold. 
Interestingly, a strongly momentum-dependent Fermi velocity can arise as 
a consequence of momentum-dependent red shift even without a direct renormalization of the band dispersion energy.
Namely, the Fermi velocity can acquire momentum-dependence 
through a momentum-dependent 
dilatation of frequency :
a quantum correction to
the frequency-dependent kinetic term of electron is translated into a renormalization of the Fermi velocity.
This momentum-dependent red shift is indeed the primary mechanism by which the Fermi velocity acquires a strong momentum dependence in the antiferromagnetic quantum critical metal in two space dimensions\cite{BORGES2023169221}.
In such cases, one can understand phenomena associated with momentum-dependent Fermi velocity as consequences of a curved momentum-spacetime geometry. 
The same physics can be 
 understood without invoking a curved momentum-spacetime, 
just as the effects of curved spacetime in general relativity can be still described within the Newtonian framework.
However, such non-geometric descriptions require introducing features that are rather arbitrary and finely tuned.
For example, through the momentum-dependent red shift, only the magnitude of the Fermi velocity is renormalized 
even if the direction is not protected by generic quantum corrections that renormalize the Fermi velocity.
In the geometric description,
 it is naturally captured as a renormalization of the temporal metric.
For this reason, we 
 adopt the geometric perspective to describe the dynamics of quasiparticles that are subject to momentum-dependent quantum corrections.

The antiferromagnetic quantum criticality is believed to play an important role in electron doped cuprates\cite{HELM}, iron pnictides\cite{HASHIMOTO}, and heavy fermion compounds\cite{PARK}. 
Although the theory in two  space dimensions becomes strongly interacting at low energies, 
it has been non-perturbatively solved in the limit that the bare nesting angle is small\cite{SCHLIEF,BORGES2023169221}, building upon earlier works on the theory
\cite{
ABANOV1,
ABANOV3,
ABANOV2,
HARTNOLL,
ABRAHAMS,
PhysRevB.87.045104,
VANUILDO,
PATEL,
VARMA2,
MAIER,
VARMA3,
MAX2,
SHOUVIK,
SUNGSIKREVIEW,
MAX1,
LIHAI2,
SCHATTNER2,
GERLACH,
LIHAI,
WANG2,
MAX1,
BERGLEDERER,
SCHATTNER2,
2022arXiv220414241L}.
While electrons at the hot spots remain coupled with critical spin fluctuations down to the zero energy limit, electrons away from the hot spots are decoupled from spin fluctuations at sufficiently low energies.
Since the crossover energy scale for the decoupling depends on momentum relative to the hot spots,
quasiparticles are renormalized with a momentum-dependent red shift,
which gives rise to a curved momentum-spacetime.
If the momentum-dependent metric is singular enough, the  emergent geometry 
can exhibit an analogous black hole horizon whose presence affects the dynamics of quasiparticles significantly.
For example, the curved momentum-spacetime geometry could directly manifests itself in the cyclotron motion of quasiparticles. 
The effect of strong correlations on cyclotron mass enhancement has been the focus 
of many earlier works\cite{
Ramshaw_2015,Custers_2003,Sebastian_2008,Sebastian_2012,PhysRevB.106.195110,https://doi.org/10.48550/arxiv.1205.3045,Doiron_Leyraud_2007,Singleton_2010,badoux2016change,Grinenko_2017,Kimata_2011}.

Here is the outline of the paper.
In Sec. \ref{sec:background}, 
we review the field-theoretical functional renormalilzatoin group study of the antiferromagnetic quantum critical metal\cite{BORGES2023169221}
 with an emphasis on the results that are needed for the present work.
In Sec. \ref{sec:afqcm2d}, we cast the theory of 
 the fully renormalized quasiparticles away from the hot spots into a theory of spinors propagating in a curved momentum-spacetime.
In Sec. \ref{sec:eom}, we compute the cyclotron period of electron as a function of the bare nesting angle of the Fermi surface.
We show that the unusual dependence of the cyclotron period on the nesting angle  is a result of 
the non-trivial evolution of the curved momentum-spacetime geometry with varying nesting angle.
We conclude with a summary and discussions in Sec. \ref{sec:con}.

\section{The antiferromagnetic quantum critical metal in 2d}
\label{sec:background}

\begin{figure}[t]
	\centering
        \begin{subfigure}[b]{0.49\linewidth}
            \centering
            \includegraphics{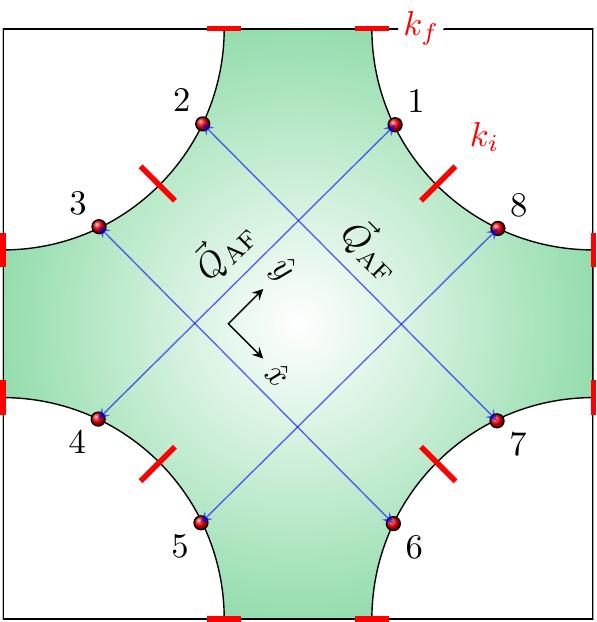}
            \caption{\label{fig:FermiSurfaceHS}}
        \end{subfigure}
        \begin{subfigure}[b]{0.49\linewidth}
            \centering
            \includegraphics[width=0.7\linewidth]{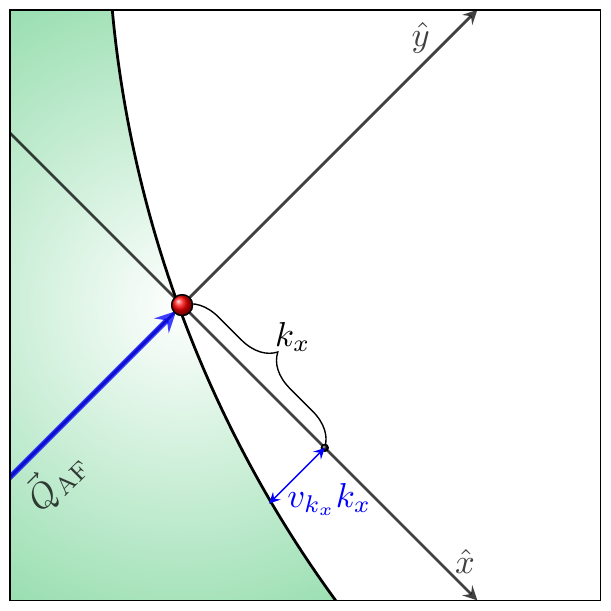}
            \caption{\label{fig:VFvectordef}}
        \end{subfigure}
            
	\caption{
For simplicity, we have omitted the superscript $\mathsf{B}$ in all the quantities in this figure. ({\color{blue}$a$}) The full Fermi surface divided into eight segments (separated by the red bars.)
Segment $1$ is bounded by $k_i$ and $k_f$,
and
other segments are related to it 
through the 
$C_4$ and reflection symmetries.
Each segment contains one hot spot denoted as red dots on the Fermi surface.
The hot spots are connected by the antiferromagnetic ordering wave vector, 
$\vec{Q}_{AF}$.
$\hat x$ ($\hat y$) is chosen to be 
perpendicular (parallel) to $\vec Q_{AF}$ 
at hot spot $1$. 
({\color{blue}$b$}) 
The Fermi surface in segment $1$ is at $v_{k_x} k_x + k_y = 0$.
Here, $v_{k_x} k_x$ represents the displacement of the Fermi surface from what the perfectly nested Fermi surface would have been.
}
\label{fig:FermiSurface}
\end{figure}
In this section, we  first review the theory of the antiferromagnetic quantum critical metal in $2+1$ dimensions,
focusing on 
the renormalized 
shape of the Fermi surface and 
the Fermi velocity 
that control 
the dynamics of low-energy quasiparticles. 
In order to capture the momentum profiles of the coupling functions across the Fermi surface, we need to consider the low-energy effective field theory that includes all gapless modes on the Fermi surface\cite{BORGES2023169221},
\begin{align}
\label{eq:LukewarmAction}
\begin{split}
&S
= \sum^{8}_{N=1}\sum^{N_c}_{\sigma =1}\sum^{N_f}_{j=1}\int\dd {\bf k}^\mathsf{B}~\psi^{\mathsf{B}\dagger}_{N,\sigma,j}({\bf k}^\mathsf{B})\left\{ik^\mathsf{B}_0 
+V^{\mathsf{B};(N)}_{F,k^\mathsf{B}_N}  e_N[\vec{k}^\mathsf{B}; v^{\mathsf{B};(N)}_{k^\mathsf{B}_N}]
\right\}\psi^\mathsf{B}_{N,\sigma,j}({\bf k}^\mathsf{B})+\\
&\frac{1}{\sqrt{N_f}}\sum^{8}_{N=1}\sum^{N_c}_{\sigma\sigma'=1}\sum^{N_f}_{j=1}
\int \dd {\bf k}^\mathsf{B}  \dd {\bf q}^\mathsf{B} ~ 
g^{\mathsf{B};(N)}_{k^\mathsf{B}_N+q^\mathsf{B}_N, k^\mathsf{B}_N}
\psi^{\mathsf{B}\dagger}_{N,\sigma',j}({\bf k}^\mathsf{B}+{\bf q}^\mathsf{B})
\Phi^\mathsf{B}_{\sigma'\sigma}({\bf q}^\mathsf{B})
\psi^\mathsf{B}_{\bar{N},\sigma,j}({\bf k}^\mathsf{B})+\\
&\frac{1}{4}\sum^{8}_{\{N_i=1\}}\sum^{N_c}_{\{\sigma_i=1\}}\sum^{N_f}_{\{j_i=1\}}
\int \prod^{4}_{i=1} 
\dd {\bf k}^\mathsf{B}_i~
\delta^\mathsf{B}_{1+2,3+4}
\lambda^{\mathsf{B};\spmqty{N_1 & N_2 \\ N_4 & N_3};\spmqty{\sigma_1 & \sigma_2 \\ \sigma_4 & \sigma_3}}_{\spmqty{k^\mathsf{B}_{1;N_1} & k^\mathsf{B}_{2;N_2} \\ k^\mathsf{B}_{4;N_4} & k^\mathsf{B}_{3;N_3}}}
\psi^{\mathsf{B}\dagger}_{N_1,\sigma_1,j_1}({\bf k}^\mathsf{B}_1)\psi^{\mathsf{B}\dagger}_{N_2,\sigma_2,j_2}({\bf k}^\mathsf{B}_2)\psi^\mathsf{B}_{N_3,\sigma_3,j_2}({\bf k}^\mathsf{B}_3)\psi^\mathsf{B}_{N_4,\sigma_4,j_1}({\bf k}^\mathsf{B}_4).
\end{split}
\end{align}
Here, we consider a Fermi surface with the $C_4$ and reflection symmetries
that supports eight hot spots labelled 
by $N=1,2,\dots,8$ 
as is shown in \fig{fig:FermiSurfaceHS}. 
The Fermi surface is divided into eight disjoint segments whose union
covers the entire Fermi surface.
Each segment, 
which contains one hot spot,
is labelled by the associated hot spot index.
$\psi^\mathsf{B}_{N,\sigma,j}({\bf k}^\mathsf{B})$ represents the electron field 
in segment $N$ with spin $\sigma=1,2,\dots, N_c$
and flavour  $j=1,2,\dots,N_f$. 
The electron is in the fundamental representations of 
spin $SU(N_c)$ and flavour $SU(N_f)$ groups.
The case that is most relevant to experiments is $N_c=2$ and $N_f=1$,
but we keep $N_c$ and $N_f$ general because
the solution obtained in the small nesting angle limit is valid for any $N_c \geq 2$ and $N_f \geq 1$. 
All quantities with superscript $\mathsf{B}$ are bare ones 
 in terms of which 
the microscopic theory is written. 
$\dd {\bf k^\mathsf{B}} \equiv \frac{ \dd k^\mathsf{B}_0 \dd k_x^\mathsf{B} \dd k_y^\mathsf{B}}{(2\pi)^3}$,
where 
${\bf k}^\mathsf{B}=(k_0^\mathsf{B},\vec{k}^\mathsf{B})=(k_0^\mathsf{B},k_x^\mathsf{B},k_y^\mathsf{B})$ denotes bare three-vector
that includes the Matsubara frequency  $k_0^\mathsf{B}$
and the two-dimensional momentum $\vec k^\mathsf{B}$ which
is measured relative to 
the hot spot in each segment. 
%
The collective antiferromagnetic spin fluctuations are represented by a bosonic field
$\Phi^\mathsf{B}({\bf q}^\mathsf{B}) = \sum_{a=1}^{N_c^2-1}\phi^{\mathsf{B};a}({\bf q}^\mathsf{B})\tau^a$, 
where $\tau^a$'s denote the $N_c\times N_c$ generators of $SU(N_c)$ with $\tr{\tau^a\tau^b}=2\delta_{ab}$
and
$\vec{q}^{~\mathsf{B}}$ is measured relative to the ordering wavevector, $\vec{Q}_{AF}$. 
We consider the case where $2 \vec Q_{AF}$ is equivalent to a reciprocal vector
and $\phi^{\mathsf{B};a}({\bf q}^\mathsf{B}) = \phi^{\mathsf{B};a}(-{\bf q}^\mathsf{B})^*$. 
The momentum conserving delta function is denoted as $\delta^\mathsf{B}_{1+2,3+4} \equiv
(2\pi)^3\delta({\bf k}^\mathsf{B}_1+{\bf k}^\mathsf{B}_2-{\bf k}^\mathsf{B}_3-{\bf k}^\mathsf{B}_4)$.
$k^\mathsf{B}_N$ represents the component of momentum that labels the Fermi surface in segment $N$,
\bqa
k^\mathsf{B}_N &=& 
\left\{ 
\begin{array}{r}
   k^\mathsf{B}_x ~~~~~ \mbox{ for $N=1,4,5,8$} \nn
   k^\mathsf{B}_y ~~~~~~\mbox{for $N=2,3,6,7$} 
 \end{array}
\right..
\eqa
$\bar{N}$ represents the hot spot connected to $N$ by the ordering vector $\vec{Q}_{AF}$; for example, 
$\Bar{N} = 4$ for $N=1$.
Although neither $\hat x$ nor $\hat y$ direction is 
perfectly parallel to the Fermi surface in general,
there is one-to-one correspondence 
between $k^\mathsf{B}_N$ and a point on the Fermi surface in each segment.
$V^{\mathsf{B};(N)}_{F,k^\mathsf{B}_N}$ is the bare momentum-dependent Fermi velocity in the direction parallel to $\vec Q_{AF}$ in segment $N$. It is the ``bare'' coupling because is the cutoff-dependent value in the action indicating the starting point where the RG flow will begin to run from.
$e_N[\vec{k}^\mathsf{B}, v^{\mathsf{B};(N)}_{k^\mathsf{B}_N}]$
specifies the Fermi surface  in segment $N$ through
\bqa
\begin{array}{llll}
 e_{1}[\vec{k}^\mathsf{B};v^{\mathsf{B};(1)}_{k^\mathsf{B}_x}] =  v^{\mathsf{B};(1)}_{k^\mathsf{B}_x} k^\mathsf{B}_x + k^\mathsf{B}_y, &
 e_{2}[\vec{k}^\mathsf{B};v^{\mathsf{B};(2)}_{k^\mathsf{B}_y}] = - v^{\mathsf{B};(2)}_{k^\mathsf{B}_y} k^\mathsf{B}_y - k^\mathsf{B}_x, \\
 e_{3}[\vec{k}^\mathsf{B};v^{\mathsf{B};(3)}_{k^\mathsf{B}_y}] = v^{\mathsf{B};(3)}_{k^\mathsf{B}_y} k^\mathsf{B}_y - k^\mathsf{B}_x, &
 e_{4}[\vec{k}^\mathsf{B};v^{\mathsf{B};(4)}_{k^\mathsf{B}_x}] =  v^{\mathsf{B};(4)}_{k^\mathsf{B}_x} k^\mathsf{B}_x - k^\mathsf{B}_y, \\
 e_{5}[\vec{k}^\mathsf{B};v^{\mathsf{B};(5)}_{k^\mathsf{B}_x}] =  -v^{\mathsf{B};(5)}_{k^\mathsf{B}_x} k^\mathsf{B}_x - k^\mathsf{B}_y, &
 e_{6}[\vec{k}^\mathsf{B};v^{\mathsf{B};(6)}_{k^\mathsf{B}_y}] =  v^{\mathsf{B};(6)}_{k^\mathsf{B}_y} k^\mathsf{B}_y + k^\mathsf{B}_x, \\
 e_{7}[\vec{k}^\mathsf{B};v^{\mathsf{B};(7)}_{k^\mathsf{B}_y}] = - v^{\mathsf{B};(7)}_{k^\mathsf{B}_y} k^\mathsf{B}_y + k^\mathsf{B}_x, &
 e_{8}[\vec{k}^\mathsf{B};v^{\mathsf{B};(8)}_{k^\mathsf{B}_x}] =  -v^{\mathsf{B};(8)}_{k^\mathsf{B}_x} k^\mathsf{B}_x + k^\mathsf{B}_y.
 \end{array}
 \label{eq:e1vk2}
\eqa
If 
$v^{\mathsf{B};(N)}_{k_{N}^\mathsf{B}}=0$, 
pairs of segments connected by $\vec Q_{AF}$ become perfectly nested. 
In general, $v^{\mathsf{B};(N)}_{k_N^\mathsf{B}}$ is non-zero and $k^\mathsf{B}$-dependent.
For this reason, we call $v^{\mathsf{B};(N)}_{k^\mathsf{B}}$ the bare momentum-dependent nesting angle.
The full Fermi velocity vector in term of $V^{\mathsf{B};(N)}_{F,k^\mathsf{B}_N}$ and $v^{\mathsf{B};(N)}_{k^\mathsf{B}_N}$ is given by
\begin{equation}
    \vec{v}^\mathsf{B}_{F}(k_x^\mathsf{B}) = V^{\mathsf{B};(1)}_{F,k_x^\mathsf{B}}\left(\frac{\partial v^{\mathsf{B};(1)}_{k_x^\mathsf{B}}}{\partial k_x^\mathsf{B}}k_x^\mathsf{B} + v^{\mathsf{B};(1)}_{k_x^\mathsf{B}}\right)\hat{x} + V^{\mathsf{B};(1)}_{F,k_x^\mathsf{B}}\hat{y},
\end{equation}
where $N=1$ (See Fig. \ref{fig:VFvectordef}.)
The Yukawa coupling denoted as $g^{\mathsf{B};(N)}_{{k'}_N^\mathsf{B}, k_{\Bar{N}}^\mathsf{B}}$ 
is a function of initial and final momenta of electrons.
Similarly, $\lambda^{\mathsf{B};\spmqty{N_1 & N_2 \\ N_4 & N_3};\spmqty{\sigma_1 & \sigma_2 \\ \sigma_4 & \sigma_3}}_{
\spmqty{k^\mathsf{B}_{1} & k^\mathsf{B}_{2} \\ k^\mathsf{B}_{4} & k^\mathsf{B}_{3}}
}
$
denotes the short-range four-fermion interaction 
labelled by momenta of electrons on the Fermi surface. 
The coupling functions in different segments are related to each other through symmetry, and
$v^{\mathsf{B};(N)}_{k^\mathsf{B}}$, 
$V^{\mathsf{B};(N)}_{F,k^\mathsf{B}}$ 
and 
$g^{\mathsf{B};(N)}_{{k'}^\mathsf{B},k^\mathsf{B}}$ 
can be written as
\bqa
\left( v^{\mathsf{B};(N)}_{k^\mathsf{B}}, ~V^{\mathsf{B};(N)}_{F,k^\mathsf{B}}, ~g^{\mathsf{B};(N)}_{{k'}^\mathsf{B},k^\mathsf{B}}  \right)
= \left\{ \begin{array}{ll}
\left( v^{\mathsf{B}}_{k^\mathsf{B}}, ~V^{\mathsf{B}}_{F,k^\mathsf{B}}, ~g^\mathsf{B}_{{k'}^\mathsf{B},k^\mathsf{B}}  \right)
, & N=1,3,4,6 \\
\left( v^{\mathsf{B}}_{-k^\mathsf{B}}, ~V^{\mathsf{B}}_{F,-k^\mathsf{B}}, ~g^\mathsf{B}_{-{k'}^\mathsf{B},-k^\mathsf{B}}  \right)
, & N=2,5,7,8 
\end{array}
\right..
\label{eq:vVgN_vVg}
\eqa
Similarly, four-fermion coupling functions 
that are mapped to each other under the symmetry are related. 
Under Gaussian scaling, in which the kinetic terms are kept invariant, the Yukawa coupling has scaling dimension $[g]=1/2$. 
The perturbative expansion is organized in powers of $g/E^{1/2}$ at energy scale $E$, and becomes uncontrolled at low energies.
Fortunately, the theory can be solved non-perturbatively in the limit that the nesting angle is small.  
The interacting low-energy fixed point obeys a new scaling relation\cite{SHOUVIK2} in which
the boson acquires an $O(1)$ anomalous dimension.
At the interacting fixed point,
the Yukawa coupling becomes marginal while the local bosonic kinetic term is irrelevant.
This is why the boson kinetic term is not included in Eq. \eqref{eq:LukewarmAction}.
The details of the non-perturbative solution can be found in Refs. \cite{SCHLIEF,BORGES2023169221}. 
In the remaining of this section, we recap the results that are needed for this paper.

Physical observables such as the correlation functions of the theory are fully encoded in the one-particle irreducible vertex functions.
We denote the vertex function of $2m$ fermions and $n$ bosons as
$\varGamma^{(2m,n)}({\bf k}_1,...,{\bf k}_{2m+n-1})$,
where ${\bf k}_i$ represents the energy-momentum vector of the $i$-th particle.
The vertex functions are defined in terms of  renormalized fields
$\psi_{N,\sigma,j}$
and $\Phi_{\sigma',\sigma}$,
which are related to the bare fields through multiplicative field renormalization factors\cite{BORGES2023169221}.
All variables without superscript $\mathsf{B}$ are renormalized quantities.
Renormalized frequency and momentum can be also related to the bare ones.
Here, we use $\vec k= \vec k^\mathsf{B}$.
However, the renormalized frequency is chosen to be related to the bare frequency through a non-linear scale transformation,
$
k_0^B = k_0 e^{-\int_{k_0}^{\Lambda} d k_0' 
z_\tau( k_0')
}
$,
where
$z_\tau( k_0')$
is the scale-dependent dynamical critical exponent
and $\Lambda$ is the scale at which the bare and renormalized frequencies coincide.
In principle, we can rescale frequency however we want, but it is convenient to choose the dynamical critical exponent such that the renormalized Fermi velocity at the hot spots is fixed to be $1$ when measured with the renormalized frequency\cite{SCHLIEF,BORGES2023169221}. 
The set of momentum and energy dependent vertex functions that control all other vertex functions is related to the coupling functions 
$\left\{
v_k, V_{F,k}, g_{k',k}, 
\lambda^{\spmqty{N_1 & N_2 \\ N_4 & N_3};\spmqty{\sigma_1 & \sigma_2 \\ \sigma_4 & \sigma_3}}_{\spmqty{k_{1;N_1} & k_{2;N_2} \\ k_{4;N_4} & k_{3;N_3}}}
\right\}$.
In this paper, we only need the momentum-dependent nesting angle $v_k$ and Fermi velocity $V_{F,k}$ defined through 
\bqa
	Re \varGamma^{(2,0)}_{1}({\bf k})\bigg|_{
 {\bf k} = \left( \mu, k_x, -v_{k_x} k_x \right)}  =  0, 
 \quad \quad
 \frac{\partial }{\partial k_y} Re 
 \varGamma^{(2,0)}_{1}({\bf k})
 \bigg|_{
 {\bf k} =\left( \mu, k_x, -v_{k_x} k_x \right)} &=& V_{F, k_{x}
 }, 
 \label{eq:VFk}
\eqa
where $\varGamma^{(2,0)}_{N}({\bf k})$ is the two-point vertex function for electron in segment $N$.
The first equation in \eq{eq:VFk} should be viewed as the defining equation for the renormalized nesting angle $v_{k}$ that determines the shape of the Fermi surface at renormalized frequency $\mu$.
The second equation defines the renormalized Fermi velocity through the vertex function evaluated on the Fermi surface.
Once the shape and Fermi velocity are fixed in segment $1$, the symmetry fix them in all other segments.
The field renormalization of electron 
and the dynamical critical exponent are chosen such that $\frac{\partial }{\partial k_0}Im \varGamma^{(2,0)}_{1}({\bf k})\bigg|_{{\bf k} =\left(\mu, k_x, -v_{k_x} k_x \right)} = i$
and $V_{F,0}=1$.

To simplify the notation,
from now on we will use $k$ and $k_x$ interchangeably. 
The evolution of $v_k$ and $V_{F,k}$ with decreasing $\mu$
defines the functional renormalization group flow\cite{BORGES2023169221}.
Below, we will use $\ell \equiv \log \Lambda/\mu$ to denote the logarithmic length scale that increases as the low-energy limit is taken,
where $\Lambda$ is a UV cutoff energy scale.
In the small nesting angle limit, 
the self-energy that determines the renormalization of $v_k$ and $V_{F,k}$ can be computed.
For the full details, 
see Appendix \ref{app:FRG}
and 
Sec. {\bf VI} of Ref. \cite{BORGES2023169221}.
In the space of coupling functions,
the theory supports an {\it interacting} fixed point\cite{SCHLIEF,BORGES2023169221} at which
the anomalous dimension of the boson is 
$O(1)$. 
As the theory flows toward the fixed point,
the coupling functions acquire non-trivial momentum dependence due to the momentum-dependent quantum corrections.
In this paper, we consider the the case in which the coupling functions are independent of momentum at the UV energy scale : $v_{k}(\ell = 0) = v_0(0), V_{F,k}(\ell =0) = 1$.
Different choices of UV couplings do not change the universal singularities of 
 the renormalized coupling functions that play the important role as we will see later.

Electrons on the Fermi surface receive quantum corrections in a momentum-dependent manner
because electrons away from the hot spots are decoupled from spin fluctuations below a momentum-dependent crossover energy scale.
The crossover scale is given by
$E^{(2L)}_k \equiv
\Lambda e^{- \ell^{(2L)}_{k}}$
(see Appendix \ref{app:FRG} for more details),
where
\begin{equation}
    \ell^{(2L)}_{k}
    = \log\left(\frac{\Lambda}{4 V_{F,k} v_k \abs{k}}\right)
    \label{eq:ellk2l}
\end{equation}
denotes the logarithmic length scale associated with the crossover.
This crossover occurs because electrons away from hot spots can not interact with with critical spin fluctuations with zero energy while staying on the Fermi surface due to a lack of the perfect nesting for $v_k \neq 0$.
Here, we emphasize that $v_k$ is small but non-zero in the non-perturbative solution. 
This momentum-dependent crossover creates two {\it momentum scales},  $k_c(\ell)$ and $k_h(\ell)$ 
determined from 
$\ell^{(2L)}_{k_c} = 0$ and $\ell^{(2L)}_{k_h} = \ell$, respectively. 
They divide the momentum space into three regions at a finite length scale $\ell$.
To the leading order in $v_k$, these mometum scales can be approximated as
$k_c \approx \frac{\Lambda}{4 v_0(0)}$ and $k_h \approx \frac{\Lambda e^{-\ell}}{4 v_0(0)}$.
In the `cold' region with $k > k_c$, electrons are too far away from the hot spot to receive significant quantum correction at energies below $\Lambda$. 
In the `lukewarm' region with $k_h(\ell) < k < k_c$,
electrons receive non-trivial quantum corrections between
 $\Lambda$ and 
 $E^{(2L)}_k $.
But, the electrons in the lukewarm region are largely decoupled 
 from spin fluctuations at energy scale $\ell$.
In the `hot' region with $k< k_h(\ell)$, electrons remain  strongly coupled with critical spin fluctuations at scale $\ell$. 
This gives rise to the following momentum profiles for the renormalized nesting angle $v_k$ and Fermi velocity $V_{F,k}$\cite{BORGES2023169221} at scale $\ell$,
\begin{align}
    v_{k} = & 
   \begin{cases}
   \FR{ \ell_0\log(\ell_0)} 
{(\ell+\ell_0)\log(\ell+\ell_0)} 
   v_0(0) 
   & 0 \leq k < k_h \\
   \FR{ \ell_0\log(\ell_0)} {(\ell^{(2L)}_k+\ell_0)\log(\ell^{(2L)}_k+\ell_0)} 
   v_0(0) 
   & k_h \leq k < k_c \\
v_0(0) 
 & k_c \leq k
   \end{cases}, \label{eq:nestinAng}\\
   V_{F,k} = &  \begin{cases}
      1 & 0\leq k < k_h \\
       \exp\Big(\sqrt{N_c^2-1}\Big(\mathrm{Ei}(\log\sqrt{\ell+\ell_0})-\mathrm{Ei}(\log\sqrt{\ell_k^{(2L)}+\ell_0})\Big)\Big) & 
       k_h \leq k < k_c \\
       \exp\Big(\sqrt{N_c^2-1}\Big(\mathrm{Ei}(\log\sqrt{\ell+\ell_0})-\mathrm{Ei}(\log\sqrt{\ell_0})\Big)\Big) & k_c \leq k
   \end{cases}.\label{eq:FermiVel}
\end{align}
Here, $\mathrm{Ei}(x)$ is the exponential integral function and
\begin{equation}
\ell_0 = \frac{\pi^2 N_c N_f}{2(N_c^2 - 1)}\frac{1}{v_0(0)\log(1/v_0(0))}
\label{eq:ellzero}
\end{equation}
represents the logarithmic length scale below which the flow of the nesting angle is negligible.
We emphasize that  the nesting angle and Fermi velocity  acquire the non-trivial momentum dependence at low energies even if they are independent of momentum at the UV scale.
The nesting angle is unchanged for $k\geq k_c$ because the Fermi surface is not renormalized far away from the hot spots.
As $k$ decreases in the lukewarm region, the nesting angle decreases logarithmically in $k$ as spin fluctuations mix electrons in  segments with opposite orientations.
The nesting angle reaches the smallest value in $k< k_h$.
The nesting angle right at the hot spot continues to decrease with increasing $\ell$.
On the other hand, the Fermi velocity is $1$ for $k \leq k_h$ at all energy scale.
As discussed earlier, this is the consequence of the frequency rescaling where the renormalized frequency is chosen such that the Fermi velocity is one at the hot spots.
The price that we pay for this choice is that the Fermi velocity increases with increasing $k$ in the lukewarm and cold regions.
The Fermi velocity in the cold region keep increasing with increasing $\ell$ despite the fact that they are not renormalized by spin fluctuations.
What really happens is that electrons in the the hot region are slowed down by spin fluctuations while electrons in the cold region are not.
In the unit of time in which the the Fermi velocity is fixed to be $1$ at the hot spots,
the electrons away from the hot spots acquire Fermi velocity that keep increasing in the low-energy limit.

As $\ell$ increases, the size of hot region shrinks as more electrons become decoupled from spin fluctuations.
In the strict zero temperature limit,
the theory develops
superconducting instabilities 
due to the run-away flow of the four-fermion coupling function\cite{BORGES2023169221}.
However,
the normal state remains stable down to an energy scale that is exponentially small in $1/\sqrt{v_0(0)}$
in the limit that $v_0(0)$ is small 
and the bare four-fermion coupling is weak.
Here, we study the dynamics of electrons at energies low enough that 
electrons are decoupled from the critical spin fluctuations  almost everywhere on the Fermi surface except for the immediate vicinity of the hot spots but high enough that the superconducting instability  is absent.
As a first step, 
we consider the dynamics of quasiparticles at zero temperature, ignoring the superconducting instability.
Later, we consider the thermal effect that arises above the superconducting transition temperature.

\section{Emergence of a curved momentum-spacetime}
\label{sec:afqcm2d}

At zero energy ($\ell = \infty$) the Fermi velocity away from the hot spots 
becomes infinite.
This represents the fact that electrons at the hot spots becomes infinitely slower than the rest of electrons.
%
While the choice of the renormalized frequency is convenient for describing the scaling behaviour of electrons at the hot spots and the critical spin fluctuations\cite{SCHLIEF}, it is not useful for describing the dynamics of electrons away from the hot spots.
For electrons away from hot spots,
it is more convenient to use the bare clock with respect to which the velocity of the cold electrons is fixed to be $1$.
We can go back to the bare unit of frequency by undoing the rescaling the frequency as
\begin{equation}
    k_0 = 
\left(    \frac{V^{(N)}_{F,k_c} }{ V^{(N)}_{F,0} } \right)
    \omega , 
    \quad \quad 
    \psi_{N,\sigma,j}({\bf k}) = 
    \left( \frac{V^{(N)}_{F,k_c} }{ V^{(N)}_{F,0} } \right)^{-1}
    \tilde{\psi}_{N,\sigma,j}(\omega,\vec{k}),
    \label{eq:frequencyrescaling}
\end{equation}
where 
we use $\omega \equiv k_0^\mathsf{B}$ for simplicity of notation, 
and the normalization of the field is chosen to keep the canonical form of the quantum effective action,
\begin{equation}
    \Gamma_{\mathrm{kin}} = \sum_{N=1}^8 \sum_{\sigma = 1}^{N_c} \sum_{j=1}^{N_f} \int  \frac{\dd\omega\dd^2 \vec k}{(2\pi)^3}\tilde{\psi}^{\dagger}_{N,\sigma,j}(\omega,\vec{k})\left\{i\omega+
     \mathcal{V}^{(N)}_{F,k_N} 
    e_N\left[\vec{k};v^{(N)}_{k_N}\right]\right\}\tilde{\psi}_{N,\sigma,j}(\omega,\vec{k})
    \label{eq:quadraticaction}.
\end{equation}
For the rest of the paper, we will use the bare frequency.
Here, $    \mathcal{V}^{(N)}_{F,k_N} = 
\left( \frac{
V^{(N)}_{F,0}
}{V^{(N)}_{F, k_c }} \right)
V^{(N)}_{F,k_N}
$
denotes the Fermi velocity measured in the  bare time.

In the low-energy limit,
$k_h$ approaches zero
and the hot region shrinks to points.
This implies that quasiparticles are well defined everywhere on the Fermi surface except at the hot spots.
The dynamics of the fully renormalized quasiparticles is described by the quadratic action\footnote{ 
Besides the quadratic action,
there also exists the four-fermion coupling 
that has been generated by the critical spin fluctuations.
However, their effects on the quasiparticle motion is sub-leading 
compared to the quantum corrections that have been  already incorporated into 
$v_k$ and $V_{F,k}$
in the limit that the nesting angle is small.
}
in \eq{eq:quadraticaction} with nesting angle $v_k$ and Fermi velocity (along $\vec{Q}_{AF}$) $\mathcal{V}_{F,k}$ given by
\begin{align}
v_{k} =  
   \begin{cases}
   \FR{\pi^2 N_c N_f}{2(N_c^2-1)}  \FR{1}{(\ell^{(2L)}_k+\ell_0)\log(\ell^{(2L)}_k+\ell_0)} & 0 \leq k < k^* \\
   v_0(0) & k^* \leq k
   \end{cases}
~~ \mbox{and} ~~  \mathcal{V}_{F,k} =  
   \begin{cases}
   e^{\sqrt{N_c^2-1}\Big(-\mathrm{Ei}(\log\sqrt{\ell_k^{(2L)}+\ell_0})+\mathrm{Ei}(\log\sqrt{\ell_0})\Big)} & 
   0 \leq k < k^* \\
   \left(\frac{k}{k_c}\right)^{\alpha_1} &  k^* < k < k_c \\
   1 & k_c \leq k
   \end{cases},
   \label{eq:nestingangleaway}
\end{align}
respectively, where  
$k^* = \frac{\Lambda e^{-\ell_0}}{4 v_0(0)}$
is the momentum scale below which the flow of the nesting angle is appreciable and
is determined from $\ell_0 = \ell^{(2L)}_{k^*}$.
It is noted that the renormalized nesting angle in \eq{eq:nestinAng} can be well approximated to be $v_0(0)$ 
for $k > k^*$
in the limit that $v_0(0)$ is small (equivalently $\ell_0 \gg 1$ ).
\begin{equation}
\alpha_1 = \frac{\sqrt{N_c - 1}}{\sqrt{\ell_0 }\log \ell_0 }
\label{eq:kstar}
\end{equation}
is the critical exponent of the Fermi velocity. 
%
As expected in Eq. \eqref{eq:nestingangleaway}, $\mathcal{V}_{F,k} =1$ for $k>k_c$ and vanishes at the hot spots. 
 There are two noteworthy features in  \eq{eq:nestingangleaway}.
First, 
in $k^* < k < k_c$,
$\mathcal{V}_{F,k}$ scales with $k$
algebraically 
while $v_{k}$  is almost constant.
This is because the quantum correction that renormalizes  $\mathcal{V}_{F,k}$  is stronger than  what renormalizes $v_k$\cite{BORGES2023169221}.
As a result, $v_k$ 
is almost momentum-independent 
except in the vicinity of the hot spot within
$k<k^*$, as expressed in 
Eq. \eqref{eq:nestingangleaway} for $v_k$.
%
Second, with virtually $k$-independent 
 $v_k$ in $k>k^*$,
both $x$ and $y$ components of Fermi velocity are renormalized in the same fashion although there is no symmetry that protects 
 the direction of the Fermi velocity.
This peculiarity arises because 
the dominant renormalization of Fermi velocity
is from the quantum correction to the frequency-dependent ($i k_0$) term of the action
in \eq{eq:LukewarmAction}\cite{BORGES2023169221}.
In other words, 
the momentum dependence of 
$\mathcal{V}_{F,k}$ 
arises
because the strength of the quantum correction 
that dilates frequency 
depends on momentum along the Fermi surface.
In the scheme that uses one global clock for the entire system, 
we are forced to  transfer 
the momentum dependence of the quantum correction 
to the field renormalization
and Fermi velocity.
While this is a perfectly legitimate picture,
what the theory is really suggesting 
is to view the momentum-dependent Fermi velocity as a 
 consequence of non-uniform temporal metric on the Fermi surface.
Here, we adopt this perspective in which
electrons have momentum-independent Fermi velocity 
in $k^* < k < k_c$
once the velocity is measured with a proper time defined with respect to a momentum-dependent metric.

\begin{figure}[t]
    \centering
    \includegraphics[width=0.8\linewidth]{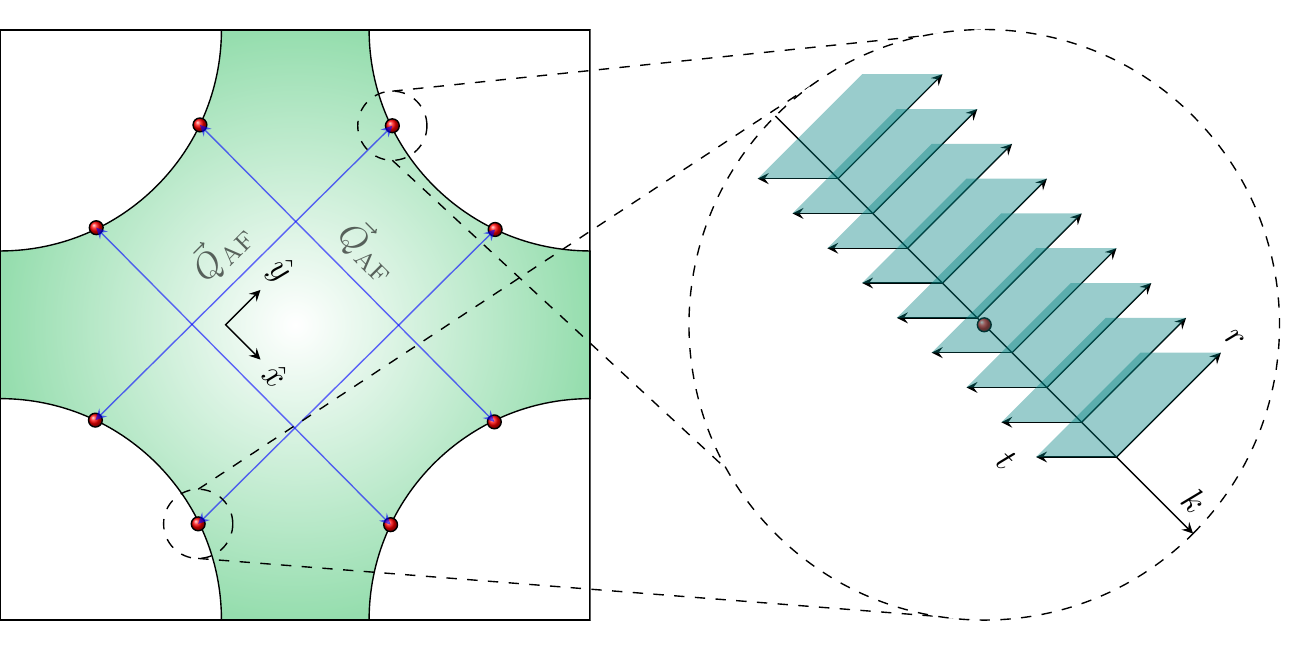}
    \caption{
    The spinor composed of the electrons in segments $1$ and $5$ is defined in 
 the hybrid spacetime $(t,r,k)$,
 where $t$ is time, 
 $r$ is space conjugate to $k_y$ and $k=k_x$. 
 At fixed $k$, electrons at hot spots 1 and 5 have the same dynamics due to the time-reversal symmetry,
 and can be naturally described by the two-component spinor $\Psi$ in Eq. \eqref{eq:spinordef}.
    }
    \label{fig:FSspacetime}
\end{figure}

\begin{figure}[th]
    \begin{subfigure}[b]{0.6\linewidth}
        \centering
        \includegraphics[width=\linewidth]{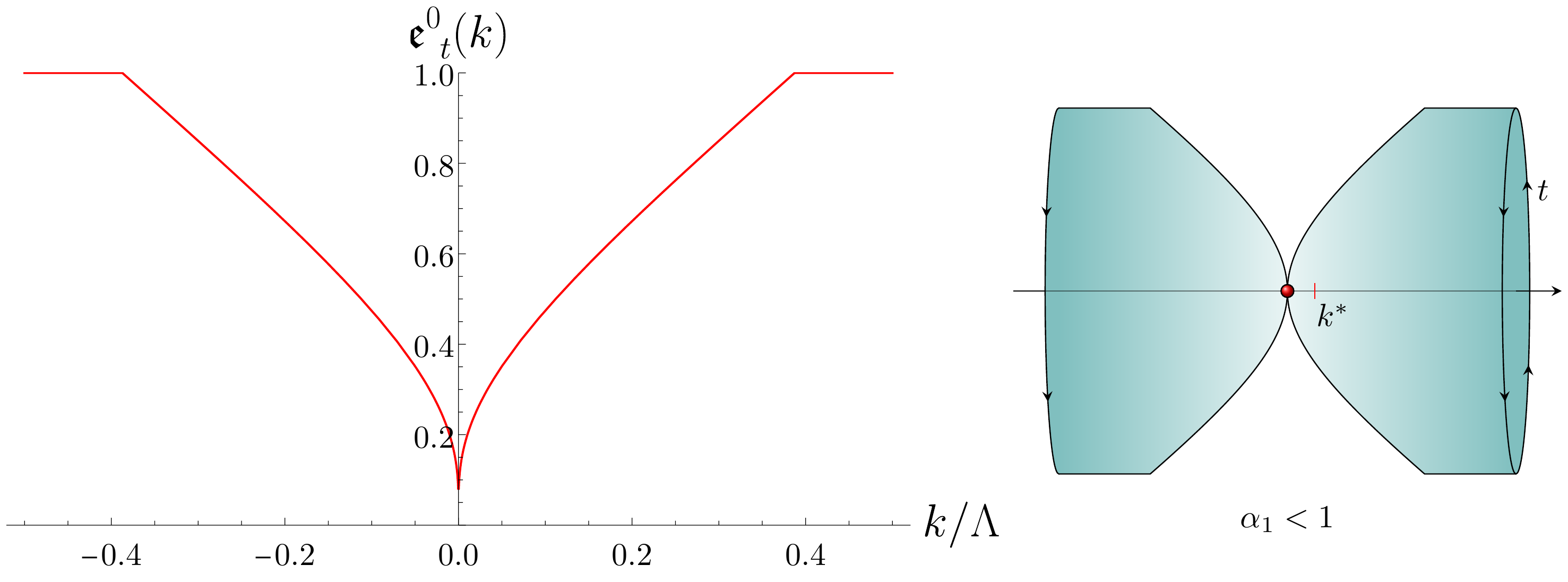}
        \caption{$\alpha_1 < 1$}
    \end{subfigure}
    \begin{subfigure}[b]{0.6\linewidth}
        \centering
        \includegraphics[width=\linewidth]{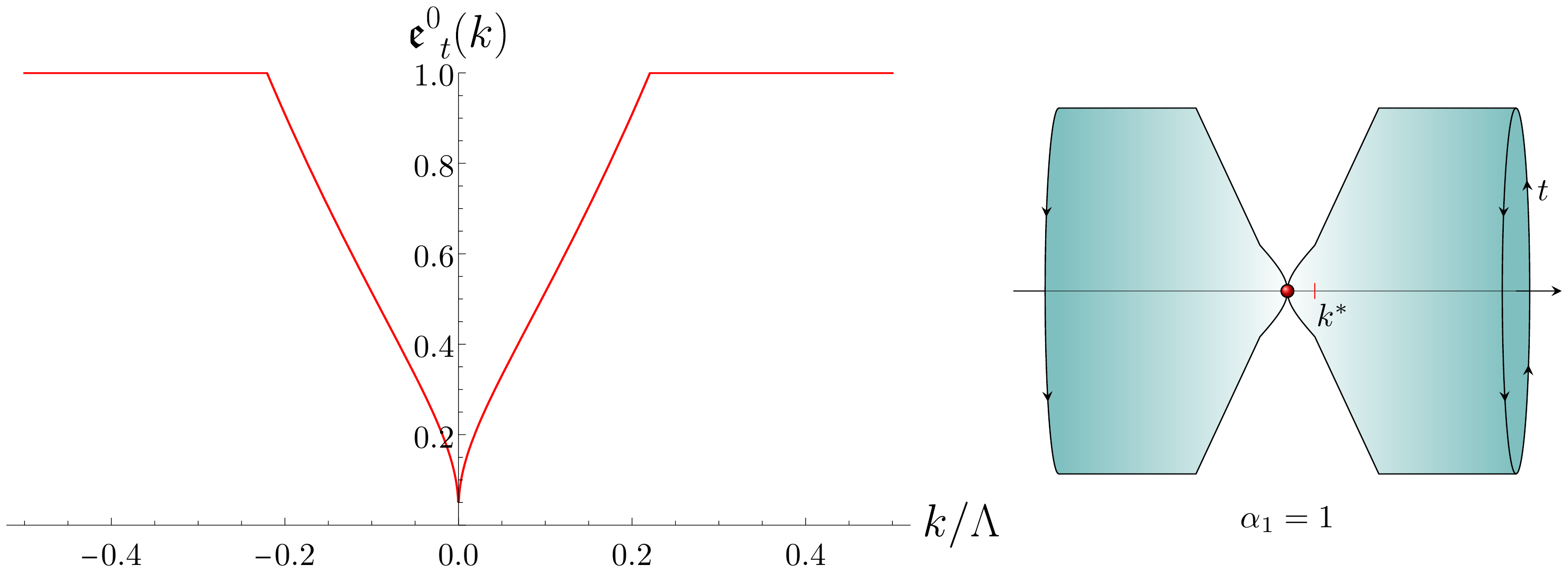}
        \caption{$\alpha_1 = 1$}
    \end{subfigure}
    \begin{subfigure}[b]{0.6\linewidth}
        \centering
        \includegraphics[width=\linewidth]{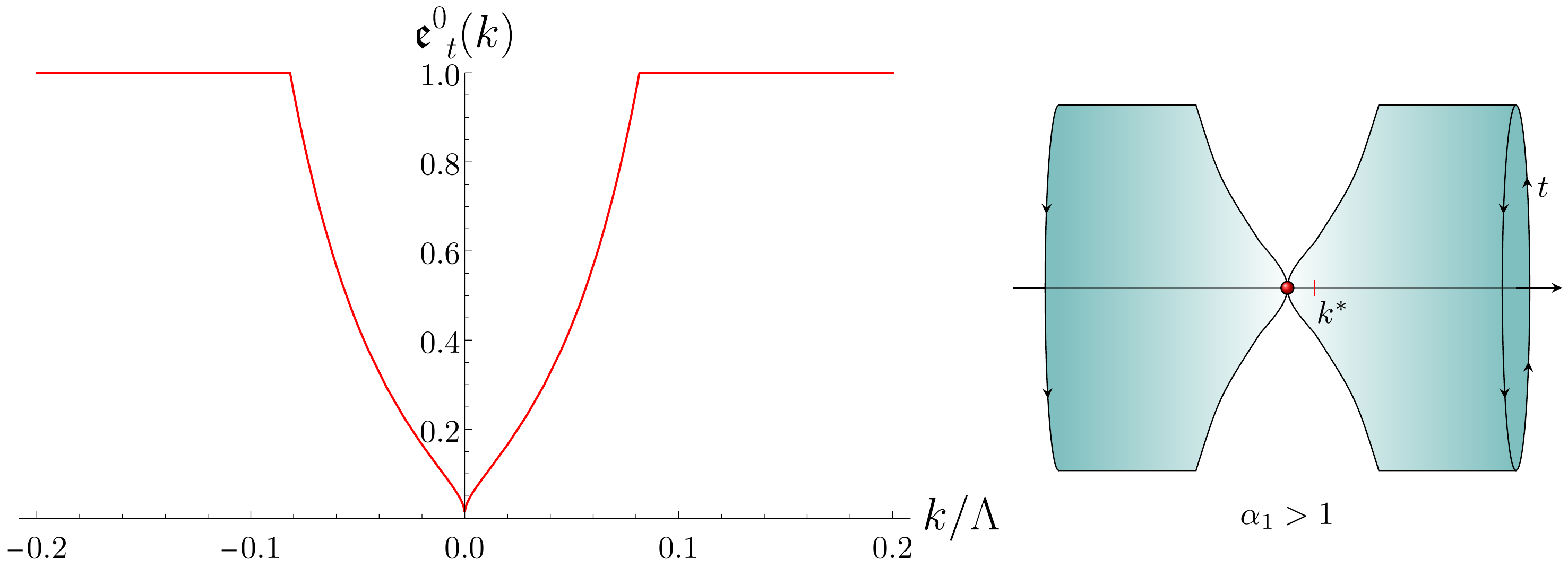}
        \caption{$\alpha_1 > 1$}
    \end{subfigure}
    \caption{
The vielbein  $\mathfrak{e}^0_{\ t}(k)$ that depends on momentum 
 along the Fermi surface determines the rate at which the proper time lapses at momentum $k$ per unit proper time of cold electrons far away from the hot spots.
The vielbein that vanishes at the hot spots represents the fact that the motion of electrons at the hot spots become infinitely slowed down compared to cold electrons.
(Left) The momentum-dependent  vielbein 
$\mathfrak{e}^0_{\ t}$ for
({\color{blue} $a$}) $\alpha_1 \approx 0.663$, 
({\color{blue} $b$}) $\alpha_1 \approx 1$ 
and ({\color{blue} $c$}) $\alpha_1 \approx 2.169$.
The choice in  ({\color{blue} $b$}) corresponds to the critical nesting angle at which the cyclotron period exhibits a logarithmic dependence on momentum  (see Eq. \eqref{eq:t12limit}). 
(Right)
 The $t-k$ slice of the hybrid spacetime for fixed $r$.
 For the purpose of illustrating the momentum-dependence of the vielbein, 
 the temporal coordinate has been compactified 
 so that the size of the circumference 
 at each $k$ represents the proper time lapsed at that momentum for every unit proper time of cold electrons.
 The circumference pinches off at the hot spots due to the infinitely large red shift at those points.
}
    \label{fig:blackHole}
\end{figure}

\begin{figure}[t]
    \centering
    \includegraphics[width=
    0.7\linewidth]{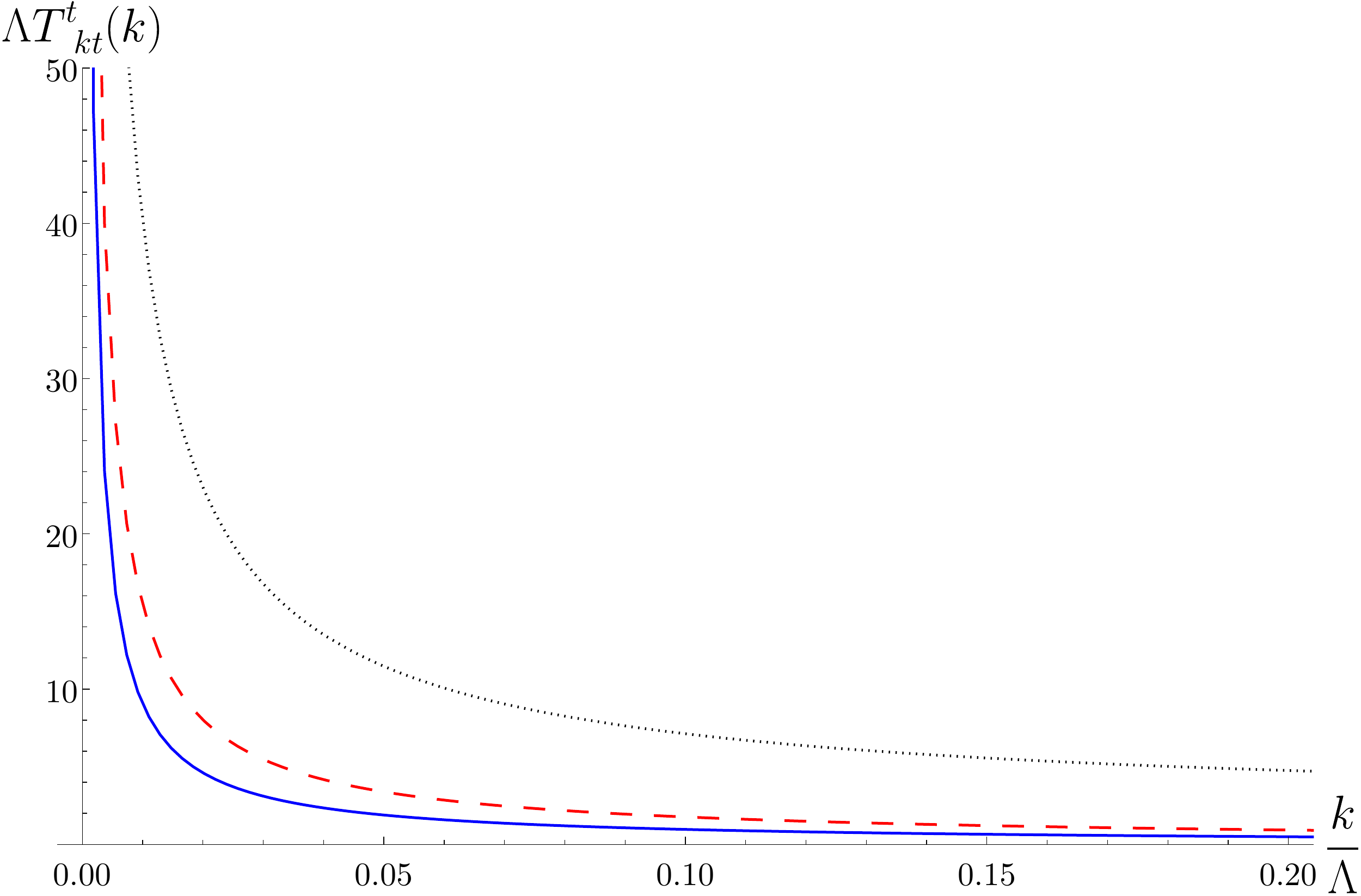}
    \caption{
    The non-zero component of torsion 
shown as a function of momentum along the Fermi surface near the hot spots.
The torsion,
as a gauge-invariant geometric quantity,
represents 
how much the momentum-spacetime in which quasiparticles propagate has been distorted from the flat one that arises in the absence of momentum-dependent quantum corrections. 
The plot is obtained from $\frac{1}{\vbt} \frac{\dd \vbt}{\dd k}$ by substituting Eq. \eqref{eq:nestingangleaway} into $\vbt$, and with the help of Eqs. \eqref{eq:ellk2l}, \eqref{eq:ellzero}, and \eqref{eq:kstar}.
The solid blue, dashed red  and dotted black curves correspond to 
$v_0(0) \approx 0.04$,
$v_0(0) \approx 0.13$ 
and
$v_0(0) \approx 1.13$, 
respectively.
This shows that the spacetime is more strongly distorted near the hot spots
and for larger bare nesting angles.
    }
    \label{fig:torsion}
\end{figure}

We formulate this geometric description by
casting  \eq{eq:quadraticaction} 
into a theory of quasiparticles
propagating in a curved  spacetime that incorporates 
the momentum-dependent metric. 
For a brief review of the fermionic action defined in curved spacetime, see Appendix \ref{app:spinors}.
For our particular case, we view the Fermi surface as a collection of $1+1$-dimensional Dirac fermions 
stacked along the direction of Fermi surface,
and combine a pair of chiral fermions at anti-podal points of the Fermi surface into a two-component Dirac spinor\cite{DENNIS}. Let us focus on segments $1$ and $5$
in this representation. Electrons in these anti-podal hot spots can be naturally paired since they have the same dynamics (See Fig. \ref{fig:FermiSurface}.) The two-component spinor is given by the Fourier transform
\begin{equation}\Psi_{\sigma, j}(t,r,k_x) 
\equiv
\int \frac{\dd \omega \dd k_y}{(2\pi)^2}
e^{i (\omega t + k_y r)}
\left[
\begin{array}{c}
\psi_{1,\sigma,j}(\omega,k_x,k_y) \\
\psi^*_{5,\sigma,j}(-\omega,-k_x,-k_y) 
\end{array}
\right].
\label{eq:spinordef}
\end{equation}
Here, the hot spot index is dropped as we focus on $N=1$ and $5$
(it is straightforward to write down the theory for other segments.)
The theory is written in the hybrid spacetime of $(t, r, k)$\cite{SHOUVIK2},  
where $t$ is time,
$r$ is the conjugate variable of $k_y$,
and
$k =k_x$ labels points on the Fermi surface in segments $1$ and $5$ (See Fig. \ref{fig:FSspacetime}.) 
From Eq. \eqref{eq:spinordef}, it is straightforward to write the action in Eq. \eqref{eq:quadraticaction} for segments $1$ and $5$ in the hybrid space of $t,r,k_x$.
The resulting action in terms of the spinors is
\begin{equation}
    \Gamma^{(1,5)}_{\mathrm{kin}} = \sum^{N_c}_{\sigma =1}\sum^{N_f}_{j=1}\int \frac{\dd k}{2\pi}\int\dd  t \dd r 
  |\vb|
  ~
    \bar \Psi_{\sigma,j}(t , r, k)
    \left\{ 
    \gamma^0
    \vb^{~t}_{0}
    D_t
    + 
    \gamma^1
    \vb^{~r}_{1}
    D_r
   \right\}
    \Psi_{\sigma,j}(t,r,k),    \label{eq:SkinGmaintemp}
\end{equation}
where $\bar \Psi = \Psi^\dagger \gamma^0$,
where
$\gamma^0 = \sigma_y$,
$\gamma^1 = \sigma_x$,
$\gamma^2=\sigma_z$
denote $2 \times 2$ gamma matrices that furnish
the two-dimensional spinor representation.
$\vb^{~\mu}_{a}$ is the inverse of the vielbein $\vb^{a}_{~\mu}$
with $a=0,1,2$
and $\mu=t,r,k$.
The vielbein determines the metric 
in the $2+1$-dimensional 
spacetime through
$g_{\mu \nu}=
\sum_{a =0}^2
\vb^{a}_{~\mu} 
\vb^{a}_{~\nu}$.
In general, the vielbein is a function of $t,r,k$,
but in our case it depends only on $k$ :
$\vb^{0}_{~t}(k) =  
\mathcal{V}_{F,k}$,
$\vb^{1}_{~r}(k) =  \vb^{2}_{~k}(k) = 1$
with all other elements being zero.
$|\vb|$ is the determinant of  $\vb^{a}_{~\mu}$.
$D_\mu = \partial_\mu + \frac{i}{2} 
\omega_{\mu,ab} 
\Sigma^{ab}
+i A_\mu$ denotes the covariant derivative,
where 
$\omega_{\mu,ab}$ is the spin connection with
$\Sigma^{ab} = \frac{i}{4}[\gamma^a, \gamma^b]$
and $A_\mu$ is the U(1) gauge field.
\eq{eq:SkinGmaintemp}
becomes equivalent to \eq{eq:quadraticaction}
for the trivial spin connection 
  $\omega_{\mu,ab}=0$
  and the gauge field 
 given by
$A_t=0$, $A_r= v_{k} k$, $A_k=0$. 
The gauge field $A_r$ gives
a $k_x$-dependent shift of momentum in the $r$ direction so that quasiparticles 
 have zero energy at $k_y = - v_{k_x} k_x$.

\eq{eq:SkinGmaintemp}
describes quasiparticles moving in a curved hybrid spacetime
with a non-trivial torsion.
It expresses the fact 
that the Fermi velocity along the direction of $\vec Q_{AF}$ is $1$ everywhere on the Fermi surface 
if the momentum-dependent proper time interval 
$\dd\tau = \vbt(k) \dd t$
is used in measuring velocity at momentum $k$.
The `apparent' variation of Fermi velocity along the Fermi surface arises  only when one chooses to probe the dynamics of quasiparticles in one fixed clock.
For an external lab  observer whose clock ticks once for every unit proper time defined in $k>k_c$,
quasiparticles appear to slow down near hot spots due to
 the momentum-dependent red-shift 
in the same way that 
a free falling object appears to undergo a slower time evolution
near the surface of a massive object with respect to the far observer due to the gravitational red shift.

The $t$-$k$ slice of the momentum-spacetime is illustrated in \fig{fig:blackHole} for the case in which the temporal direction is compact.
In $k \gg k_c$, the proper length of the temporal direction remains equal to the periodicity of $t$ since $\mathfrak{e}^0_{\ t} = 1$.
In $k^* < k < k_c$, the proper length scales with $k$ algebraically with exponent $\alpha_1$ ($\mathfrak{e}^0_{\ t} \sim k^{\alpha_1}$). 
Hence, the proper time lapses slower as the hot spot is approached.
In $k < k^*$, the power-law decay is replaced with a slower decay due to a flow of the nesting angle as is shown in Eq.\eqref{eq:nestingangleaway}.

With the vielbein and spin connection fully fixed by the renormalized coupling functions, Cartan's structure equation determines the torsion of the spacetime to be 
 $T^t=
\frac{1}{\vbt} 
\frac{\dd \vbt}{\dd k}
\dd k \wedge \dd t$, $T^r=T^k=0$.
The torsion measures the failure of closure when each of two vector is parallel transported along the other vector.
The non-zero component of the torsion indicates a non-trivial structure of the hybrid spacetime.
For the present hybrid spacetime,
the torsion diverges in the $k \rightarrow 0$ limit as is shown in  \fig{fig:torsion}. 
In principle, the torsion can be measured from the rate at which the red shift varies along the Fermi surface.
However, it is not clear how the  momentum-dependent torsion can be directly measured from an experimental probe that is local in momentum space.
Here, we consider a physical observable that probes the global aspect of the distorted momentum-spacetime, which is sensitive to the torsion yet much easier to measure experimentally.



%

\section{
Cyclotron motion of quasiparticles 
in the curved momentum-spacetime
}
\label{sec:eom}

\begin{figure}[t]
\centering
\includegraphics[width=0.9\linewidth]{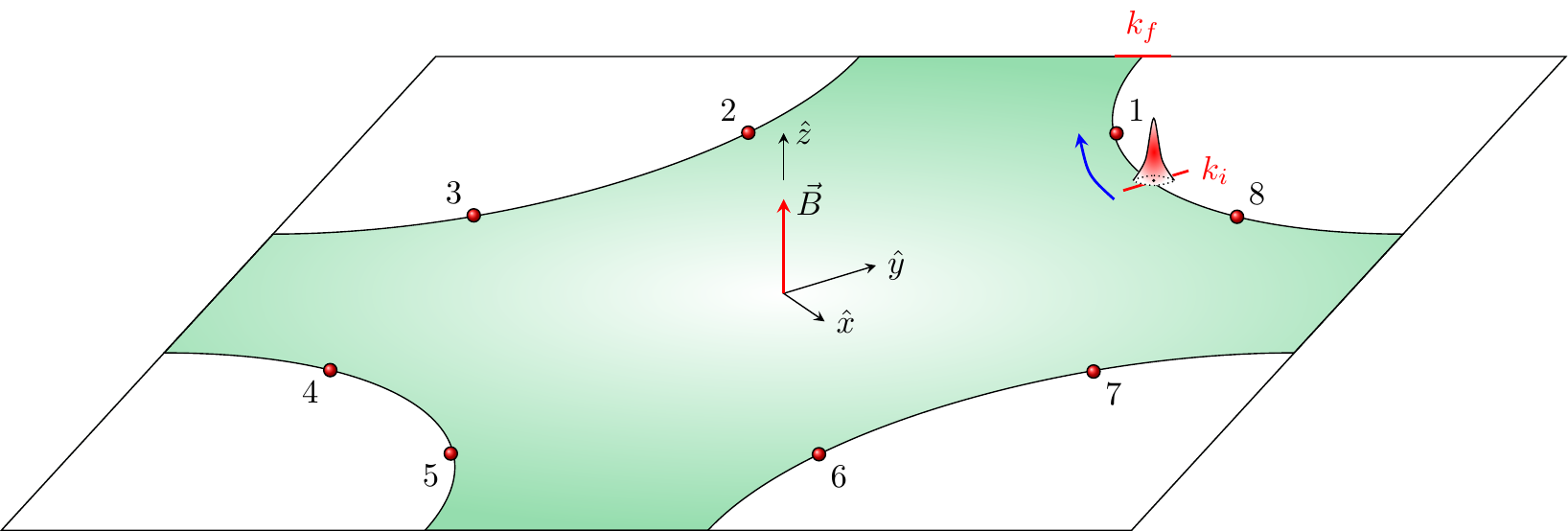}
\caption{
Initial setup of the quasiparticle wavepacket. 
The initial wavepacket of a quasiparticle is placed at the boundary between segments $1$ and $8$ ($k_i$).
In the presence of magnetic field applied in the $z$ direction, the wavepacket moves along the Fermi surface towards hot spot 1. 
As it approaches hot spot $1$, it slows down due to the momentum-dependent red shift.
}
\label{fig:setup}
\end{figure}

In this section, 
we examine how the curved momentum-spacetime affects 
the dynamics of quasiparticles
by computing 
the cyclotron period of electron in the presence of magnetic field\cite{Stanescu_2008,senthil2014mass}.
Due to the $C_4$ symmetry and the reflection symmetry around the boundary between segments, 
the cyclotron period 
at bare nesting angle $v_0$\footnote{
In this section,
we use 
$v_0$
and $v_0(0)$
interchangeably
for the momentum-independent bare nesting angle.
} 
is eight times the time it takes for a quasiparticle to traverse segment $1$ :
$T(v_0) = 8 
\left[
T(k_i,0;v_0)
+
T(0,k_f;v_0) 
\right]
$,
where 
$T(k_i,0;v_0)$ denotes
the time that it takes for a quasiparticle to traverse from 
the boundary between segments $1$ and $8$ to hot spot $1$,
and $T(0,k_f;v_0)$, 
from hot spot $1$ to the Brillouin zone boundary between segments $1$ and $6$
(see \fig{fig:FermiSurface}).
The setup is depicted in Fig. \ref{fig:setup}.  
For simplicity, we 
assume 
$T(0,k_f;v_0)=T(k_i,0;v_0)$ 
and focus on the computation of 
$T(k_i,0;v_0)$ here\footnote{
In general, 
$T(0,k_f;v_0) \neq T(k_i,0;v_0)$  because there is no reflection symmetry around the hot spots.
However, 
the computation of  $T(0,k_f;v_0)$ is exactly parallel to that of  $T(k_i,0;v_0)$. 
This is because the singular part of the quantum corrections are 
 symmetric around the hot spots\cite{BORGES2023169221}.
}.

%
In the zero temperature limit, well-defined quasiparticles exist away from the hot spots
and we can use the semi-classical description of their dynamics.
Strictly speaking, electrons right at the hot spots are not described by quasiparticle even at zero temperature.
However, the hot spots are a set of measure zero on the Fermi surface and do not affect the cyclotron period.
For a quasiparticle localized at momentum $\vec k$, its time evolution is entirely determined by the kinetic action at that momentum.
According to \eq{eq:SkinGmaintemp}, the quasiparticle has Fermi velocity
given by 
\bqa
\vec{v}_{F}(k) =
\left(
k\frac{\partial v_k}{\partial k} +v_k \right)\hat{x}
+\hat{y} 
\eqa
when the Fermi velocity is measured with respect to the proper time 
$\tau$ defined at that momentum through $\dd\tau = \mathfrak{e}^{0}_{\ t} \dd t$,
where $t$ represents the time associated with the bare frequency $\omega$.
Since the equation of quasiparticle at momentum k is entirely determined from the Fermi velocity 
and the proper time 
defined at that momentum,
the equation of motion for the the wavepacket of a quasiparticle is given by the standard equation of motion\cite{ASHCROFT},
\begin{equation}
\frac{\dd\vec{r}}{\dd \tau} = 
\vec{v}_{F}(k),
~~~~
\frac{\dd\vec{k}}{\dd \tau} =  -e ~ 
\vec{v}_{F}(k) \times \vec B,
\label{eq:dotprofiletau}
\end{equation}
when written in terms of the local proper time coordinate $\tau$,
where $\vec B = B_0 \hat z$ is the magnetic field applied along the $\hat z$ direction.
$B_0$ is assumed to be weak so that it does not affect the renormalized coupling functions.
To measure the cyclotron period in the lab frame, however,
we need to recast the equations of motion in the bare time $t$,
\begin{equation}
\frac{1}{\vbt}
\frac{\dd\vec{r}}{\dd t} = 
\left(k\frac{\partial v_k}{\partial k}+v_k\right)\hat{x}+\hat{y},
~~~~
\frac{1}{\vbt}
\frac{\dd\vec{k}}{\dd t} =  -e B_0 \left[\hat{x}-\left(k\frac{\partial v_k}{\partial k}+v_k\right)\hat{y}\right],
\label{eq:dotprofile}
\end{equation}
where the effects of the curved spacetime are captured by the vielbein $\mathfrak{e}^{0}_{\ t}$.
We consider quasiparticles on the Fermi surface, which allows us to focus on the equation of motion for the momentum along the Fermi surface.

Now let us proceed with the solutions. For $k_c < k_i$ ($k_i > \frac{\Lambda}{4v_0}$),
there exists a region of Fermi surface near the zone boundary where the renormalization from spin fluctuations is negligible at energies below UV cutoff $\Lambda$.
In this case,
$T(k_i,0;v_0)$
can be written as the sum of three intervals,
$T(k_i,0;v_0)=
T(k_i,k_c;v_0) 
+ T(k_c, k^*;v_0)
+T(k^*, 0;v_0)$.
$T(k_i,k_c;v_0)$ denotes the time that the quasiparticle spends in the region where the quantum correction from spin fluctuations is negligible 
and the hybrid spacetime is  almost flat.
$T(k_c, k^*;v_0)$ arises from the region with algebraically decaying $\vbt(k)$.
Finally,
$T(k^*, 0;v_0)$ denotes the time that the quasiparticle spends in the very vicinity of the hot spot where the flow of the nesting angle modifies the spacetime geometry from the algebraic form.
In the following, we compute each time interval one by one.

\begin{figure}[t]
    \centering
    \includegraphics[width=0.75\linewidth]{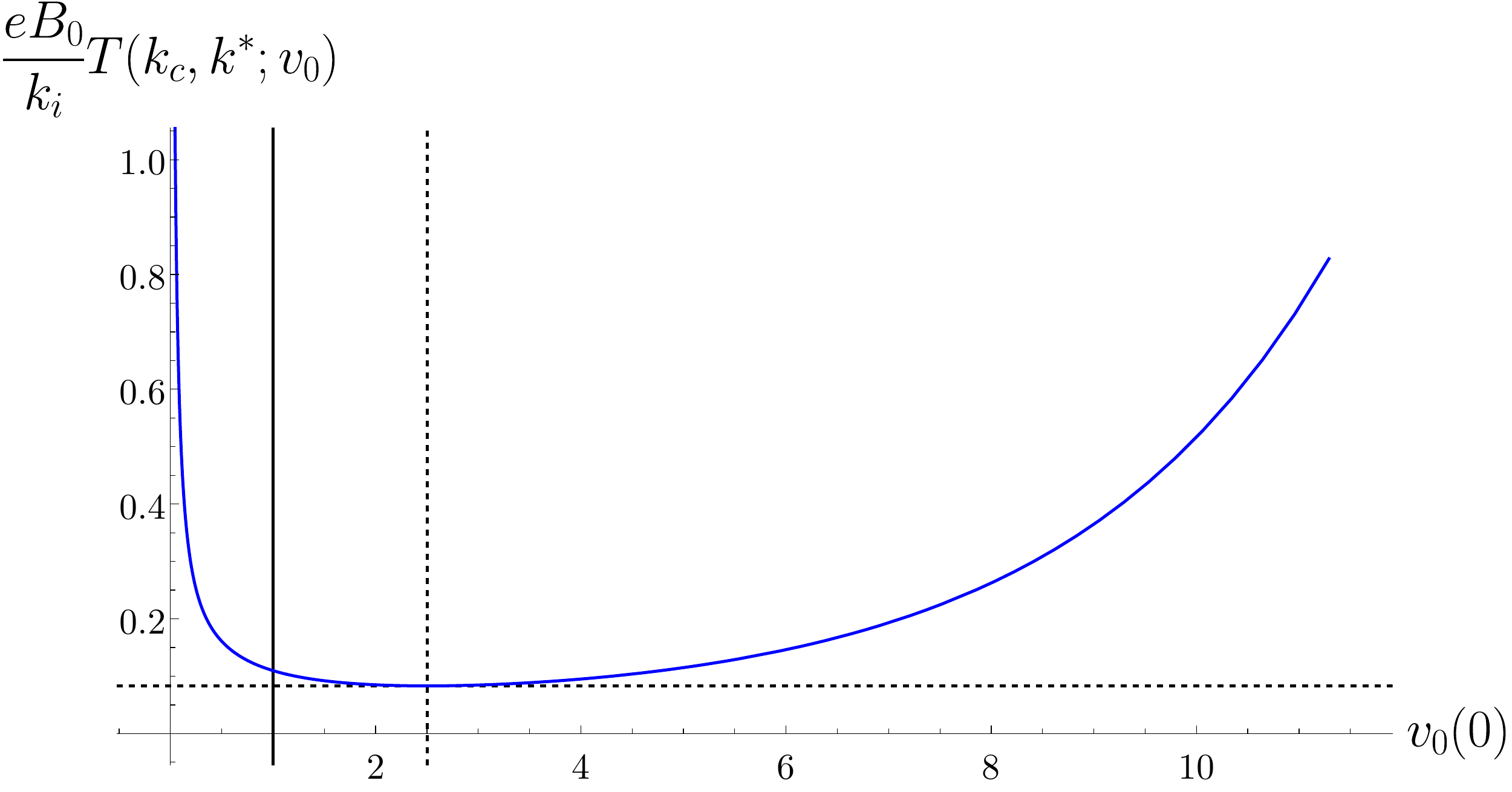}
	\caption{
	The time that it takes for a quasiparticle to traverse from $k_c$ to $k^*$ plotted in the unit of $k_i/(eB_0)$ as a function of the bare nesting angle from $v_0 \approx 0.04$ to $v_0 \approx 11$
for the choice $k_i/\Lambda=6$.
The solid vertical line denotes the nesting angle 
($v_0 \approx 1.13$)
at which $\alpha_1 = 1$. 
The dashed lines mark the minimum of 
$T(k_c,k_i;v_0)$. 
The non-monotonic behaviour of 
$T(k_c,k_i;v_0)$ 
arises from 
the interplay between two effects :
with increasing nesting angle, the size of the lukewarm region shrinks but the intensity of the red shift induced by quantum corrections is increased. 
}
	\label{fig:T12plot}
\end{figure}
In the cold region with $k_c < k < k_i$, 
$\vbt \approx 1$ (see the third line for $\mathcal{V}_{F,k}$ in Eq. \eqref{eq:nestingangleaway}).
From 
$\frac{\dd\vec{r}}{\dd t} =  v_{0}(0)\hat{x}+\hat{y}$,
$\frac{\dd\vec{k}}{\dd t} =  -e B_0\left(\hat{x}- v_0(0)\hat{y}\right)$,
one readily obtains
\begin{equation}
    T(k_i,k_c;v_0) = \frac{k_i-k_c}{e B_0}.
    \label{eq:timeitocold}
\end{equation}
In $k^* < k < k_c$, 
the nesting angle can be  still regarded as momentum-independent
while the vielbein decays as
$\vbt =  \left(k/k_c \right)^{\alpha_1}$, following the second line in Eq. \eqref{eq:nestingangleaway},
where the exponent $\alpha_1$ is determined 
from the bare nesting angle through Eqs. 
(\ref{eq:ellzero})
and
(\ref{eq:kstar}).
The equations of motion become
\begin{equation}
\frac{\dd\vec{r}}{\dd t} =   \left(\frac{k(t)}{k_c}\right)^{\alpha_1}\left(v_{0}(0)\hat{x}+\hat{y}\right),
~~~~~~
\frac{\dd\vec{k}}{\dd t} =  -e B_0\left(\frac{k(t)}{k_c}\right)^{\alpha_1}\left(\hat{x}- v_0(0)\hat{y}\right),
\label{eq:dotlowe}
\end{equation}
where $k(t)$ denotes the $x$-component of $\vec k(t)$.
Integrating
$\frac{\dd k(t)}{\dd t} = - e B_0 \left(\frac{k(t)}{k_c}\right)^{\alpha_1}$
from $k_c$ to $k^*$,
we obtain
\begin{equation}
    T(k_c,k^*;v_0) = \frac{k_c^{\alpha_1}}{e B_0}\int^{k_c}_{k^*} \frac{\dd k'}{k'^{\alpha_1}}
= 
\frac{k_c}{(1-\alpha_1)e B_0}
\left[1 - \left(\frac{k^*}{k_c}\right)^{1-\alpha_1}\right].
\label{eq:timecoldtostar}
\end{equation}
$T(k_c,k^*;v_0)$ is plotted as a function of $v_0(0)$
in Fig. \ref{fig:T12plot}. This plot is obtained by substituting the expressions in Eqs. \eqref{eq:kstar} and \eqref{eq:ellzero} in Eq. \eqref{eq:timecoldtostar}.
For small nesting angle, $T(k_c,k^*;v_0)$ rapidly decreases with increasing $v_0(0)$.
This is because
the range of lukewarm region decreases with increasing nesting angle for a fixed $\Lambda$
(see \eq{eq:crossovermscales} and \fig{fig:IRscales}).
Furthermore, 
at larger nesting angles,
even those electrons in the lukewarm region 
decouple from spin fluctuations 
at higher energy scales.
Remarkably, 
$T(k_c,k^*;v_0)$ 
bounces back
as $v_0(0)$ increases  further.
This non-monotonic behaviour is due to a competing effect that an increasing nesting angle has.
At larger nesting angles,
a reduction in the density of states of low-energy particle-hole excitations weakens the screening of interaction\cite{SCHLIEF}. 
This makes the quantum-correction-induced red shift stronger for electrons close to the hot spots.

As the nesting angle increases, 
the portion of Fermi surface affected by spin fluctuations shrinks while electrons close to the hot spots are more significantly renormalized.
The disparity in the strength of quantum correction in different parts of Fermi surface 
causes a more strongly  curved spacetime at a large nesting angle.
This is also reflected in the torsion that increases with increasing nesting angle as is shown in 
    \fig{fig:torsion}.
The metric that becomes more singular at the hot spots with increasing nesting angle creates a possibility of realizing an analogous black hole horizon in momentum space.
As $\alpha_1$ approaches $1$,
the prefactor  
$\frac{k_c}{(1-\alpha_1)e B_0}$
in \eq{eq:timecoldtostar} diverges
and
$T(k_c,k^*;v_0)$ becomes
\begin{equation}
 \lim_{\alpha_1 \to 1}
    T(k_c,k^*;v_0) = \frac{k_c}{eB_0}\log\left(\frac{k_c}{k^*}\right).
    \label{eq:t12limit}
\end{equation}
The leading small-angle expansion predicts that 
$\alpha_1$ becomes $1$  at $v_0 \approx  1.13$ for $N_c = 2$ and  $N_f = 1$. 
Even though the small $v_0$ expansion is not valid for theories with $v_0 \sim 1$, 
here we proceed with the assumption that the qualitative feature of the theory remains unchanged even at nesting angles that are not so small\cite{2022arXiv220414241L}.
In this case, 
there may well be a 
critical nesting angle at which $\alpha_1$ becomes $1$
even if the actual critical value of $v_0$ 
 differs from what is predicted from the small-$v_0$  expansion.
The way the time interval depends on $k^*$
in \eq{eq:t12limit}
is reminiscent of the logarithmic divergence in the time needed for a free-falling object to reach the horizon of the Schwarzschild black hole as measured by an asymptotic observer. This behaviour is analogous; nevertheless, it is not a coincidence. 
It is a consequence of the fact that at $\alpha_1=1$ 
the metric of a $t$ and $k$ slice 
of the hybrid spacetime in $k>k^*$
is conformally equivalent to that of the Schwazschild black hole outside the horizon. 
For a review of the Schwarzchild metric and the time needed for a free-falling object to reach the horizon, see Appendix \ref{app:SchwarzschildBH}.
If $k^*$ was zero, 
the cyclotron period would diverge 
and a quasiparticle would not be able to go through the hot spot
for $\alpha_1 \geq 1$.
In our case, the divergence is cut off by $k^*$ because the momentum-spacetime geometry
is modified
from that of the Schwarschild horizon for
$k< k^*$ due to the flow of the nesting angle
\footnote{
For this reason, the geometry that emerges in the $\alpha_1=1$ limit is more a fuzzball\cite{Mathur:2005zp} or a firewall\cite{Almheiri:2013aa} than a horizon with no `drama'.
The fact that the horizon is a special place is also seen from the fact that the torsion diverges at the hot spots (see  \fig{fig:torsion}). 
Another difference from the Schwarzschild horizon is that there is no interior of the black hole in our analogous horizon.
}.

\begin{figure}[t]
\centering
\includegraphics[width=0.75\linewidth]{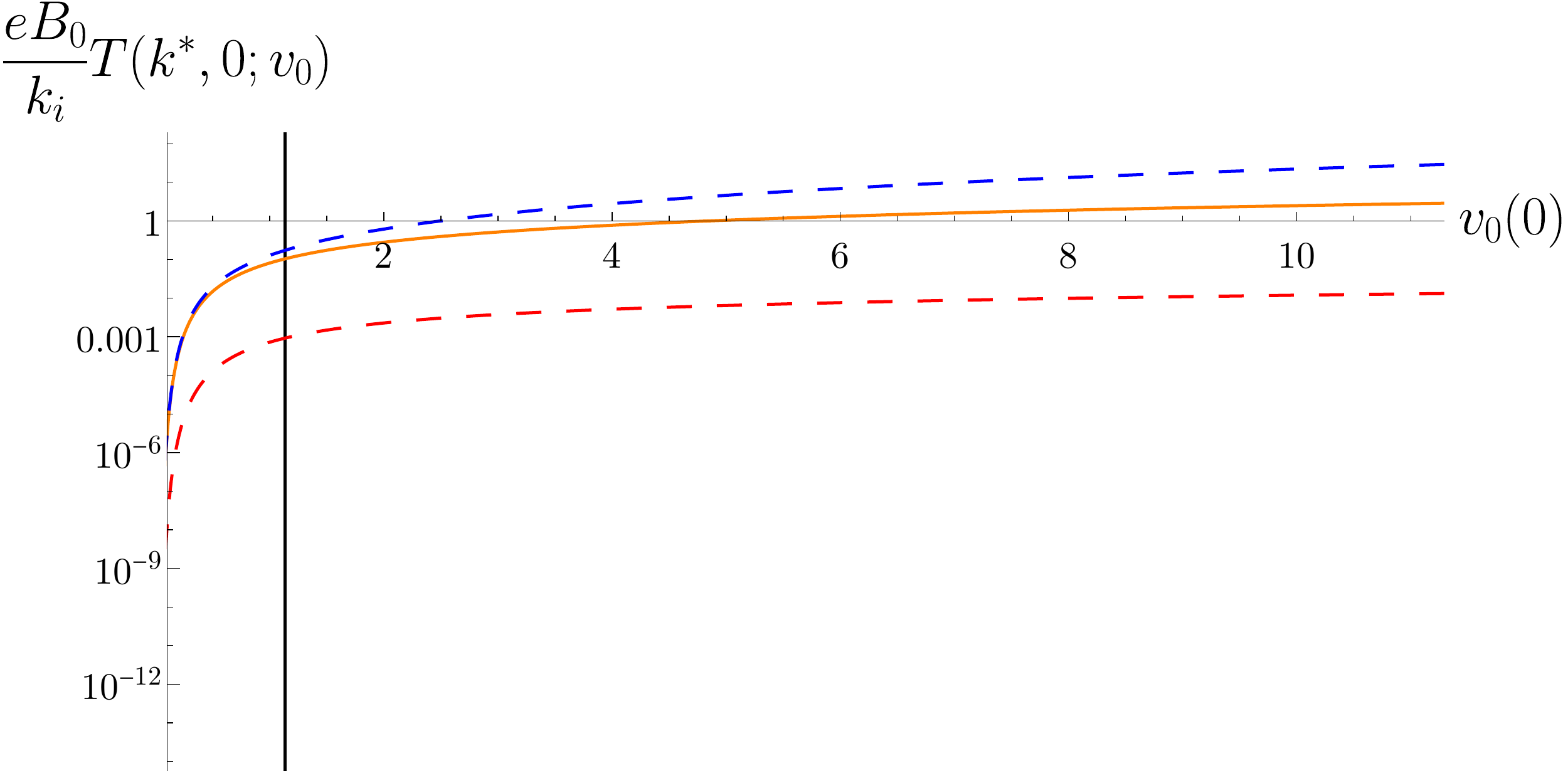}
\caption{
The solid curve represents	
$T(k^*,0;v_0)$ 
plotted as a function of  $v_0(0)$. 
Two dashed lines that sandwich the solid curve are upper and lower bounds whose expressions can be obtained analytically (see text), which shows that $T(k^*,0;v_0)$ is finite at all values of $v_0(0)$.
The vertical line marks the nesting angle at which $\alpha_1 = 1$. 
}
\label{fig:T3HS}
\end{figure}
\begin{figure}[t]
\centering
\includegraphics[width=0.75\linewidth]{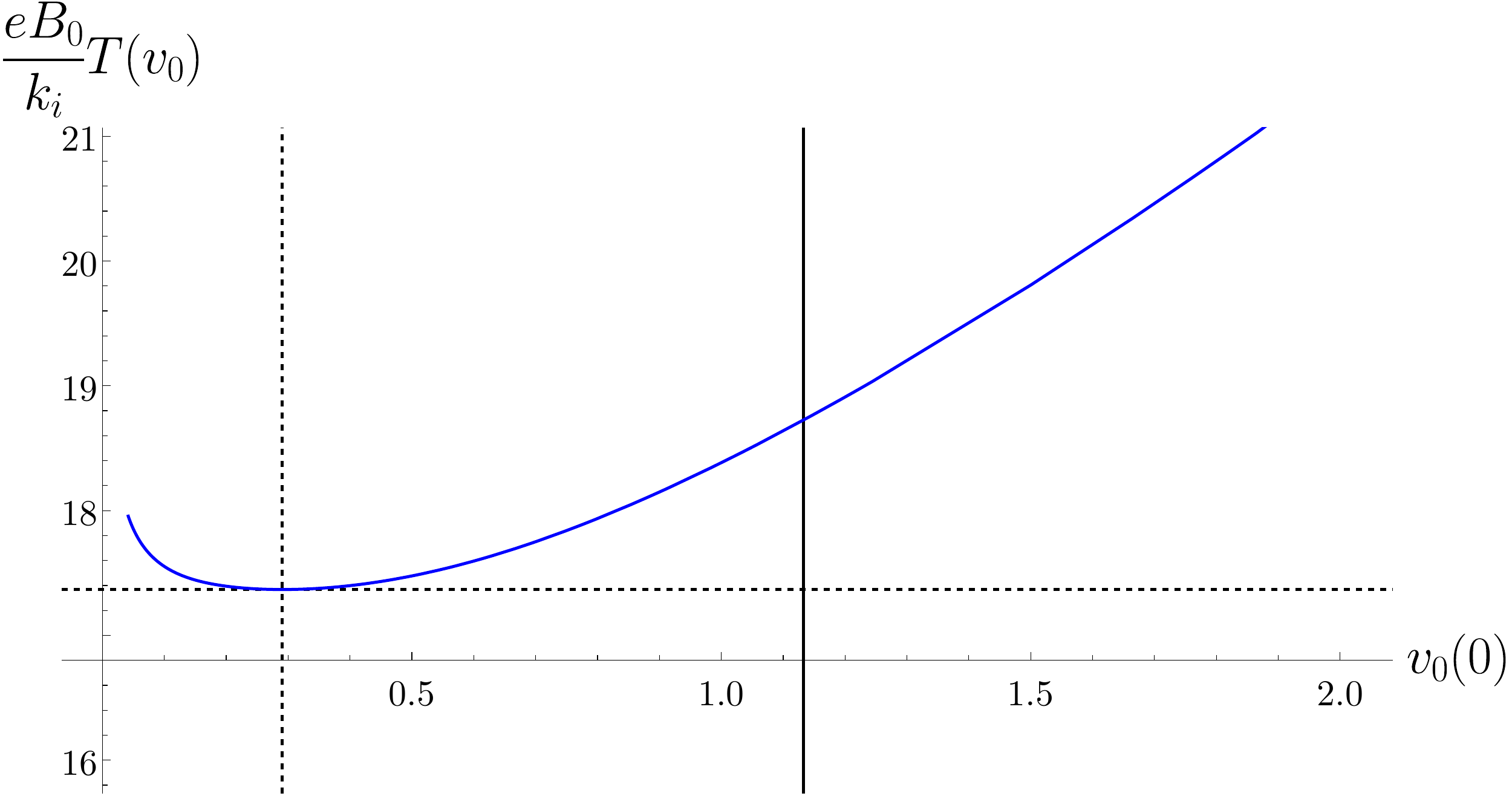}
\caption{
The cyclotron period
$T(v_0)$ plotted as a function of the bare nesting angle
for $k_i/\Lambda=6$.
This plot is obtained by adding the times computed for Figs. \ref{fig:T12plot} and \ref{fig:T3HS}.
The solid vertical line denotes the nesting angle with $\alpha_1 = 1$,
and the dashed lines mark the minimum of $T(v_0)$. 
}
\label{fig:Tfull}
\end{figure}
For $0 < k < k^*$, the electrons stay coupled with spin fluctuations at energy scales that are low enough that one has to consider the flow of the nesting angle.
This modifies the temporal vielbein from the algebraic form to a `super-logarithmic' form\footnote{
For this,  we use 
$\Ei\left(X\right) \approx \frac{e^{X}}{X}$
for $ X \gg 1$.
} in the first line of Eq. \eqref{eq:nestingangleaway},
$  \vbt
    \approx \frac{1}{\nu(\ell_0)}\exp\left(-\sqrt{N_c^2-1}\frac{\sqrt{\ell^{(2L)}_k+\ell_0}}{\log\sqrt{\ell^{(2L)}_k+\ell_0}}\right)$,
where
$\nu(\ell_0)=\exp\left\lbrace-\sqrt{N_c^2-1}\Ei\left(\log\sqrt{\ell_0}\right)\right\rbrace$.
Because the nesting angle decreases in the vicinity of the hot spot, 
the quantum correction becomes weaker.
As $k$ approaches $0$,
$\vbt$ decreases to zero 
only as $e^{-\sqrt{ \log 1/k}}$,
which is slower than any power-law.
This results in the time interval that remains finite even for $\alpha_1 \geq 1$,
$ T(k^*,0;v_0) = \frac{\nu(\ell_0)}{e B_0} \int_0^{k^*}  \exp\left(\sqrt{N_c^2-1}\frac{\sqrt{\ell^{(2L)}_{k'}+\ell_0}}{\log\sqrt{\ell^{(2L)}_{k'}+\ell_0}}\right)\dd k'$.
Substitution $s = \sqrt{\ell^{(2L)}_{k'} + \ell_0}$ yields
\begin{equation}
    T(k^*,0;v_0) = \frac{\Lambda}{2 v_0(0)}\frac{\nu(\ell_0) e^{\ell_0}}{e B_0} \int_{\sqrt{2\ell_0}}^{\infty} s \exp\left(-s^2+\sqrt{N_c^2-1}\frac{s}{\log s}\right)\dd s.
    \label{eq:T3}
\end{equation}
It is hard to evaluate Eq. \eqref{eq:T3} exactly.
But, we can bound it as
$T^{(\mathrm{l})}(k^*,0;v_0) < T(k^*,0;v_0) < T^{(\mathrm{u})}(k^*,0;v_0)$,
where the lower bound is obtained by dropping $s/\log s$ in the exponent on the integrand of \eq{eq:T3},
\begin{equation}
T^{(\mathrm{l})}(k^*,0;v_0) = \frac{\nu(\ell_0)}{e B_0}k^*,
\label{eq:T3lbound}
\end{equation}
and the upper bound is obtained by using
$\log s \approx \log\sqrt{2\ell_0}$,
\begin{equation}
T^{(\mathrm{u})}(k^*,0;v_0) = \frac{\nu(\ell_0)}{e B_0}k^* \left[e^{\sqrt{N_c^2-1}\frac{\sqrt{2\ell_0}}{\log\sqrt{2\ell_0}}} + \sqrt{\frac{\pi(N_c^2-1)}{4}}e^{\frac{N_c^2-1}{4\log^2\sqrt{2\ell_0}}+2\ell_0}\frac{ 1 + \mathrm{erf}\left(\frac{\sqrt{N_c^2-1}}{2\log\sqrt{2\ell_0}} - \sqrt{2\ell_0}\right)}{\log\sqrt{2\ell_0}}\right].
\label{eq:T3ubound}
\end{equation}
Here,
$\mathrm{erf}(z) = \frac{2}{\sqrt{\pi}}\int_0^z e^{-t^2} \dd t$
is the error function.
Fig. \ref{fig:T3HS} shows 
$T(k^*,0;v_0)$, $T^{(\text{u})}(k^*,0;v_0)$ and $T^{(\text{l})}(k^*,0;v_0)$ as functions of $v_0(0)$. 
The upper and lower bounds are direct plots of 
Eqs. \eqref{eq:T3lbound} and \eqref{eq:T3ubound}.
The solid line is obtained by 
numerically evaluating 
Eq. \eqref{eq:T3}. 
This shows that 
$T(k^*,0,v_0)$ 
is finite even for $v_0$ with $\alpha_1 \geq 1$.
$T(k^*,0,v_0)$  decreases  with decreasing nesting angle.
Especially, the sharp drop of 
$T(k^*,0,v_0)$ 
in the small $v_0$ limit is due to the decrease of 
$k^* = \frac{\Lambda e^{-\ell_0}}{4 v_0(0)}$ with decreasing $v_0$ (see Eq. \eqref{eq:ellzero}).
The cyclotron period given by the sum of time lapse in each segment,
$  T(v_0) =  16 \left[
T(k_i,k_c;v_0)  +
T(k_c,k^*;v_0) + T(k^*,0;v_0) \right]$.
$T(v_0)$ is plotted in Fig. \ref{fig:Tfull}.
The non-monotonic behaviour of $T(v_0)$ as a function of $v_0$ is attributed to the non-monotonicity of
$T(k_c,k^*;v_0)$ created by
the increasing disparity in the strength of red shift across the Fermi surface with increasing $v_0$.

So far, we have consider the zero-temperature limit in which the entire Fermi surface supports coherent quasiparticles except at the hot spots.
Since electrons are decoupled from spin fluctuations 
away from the hot spots 
 and
all quantum effects that renormalize electrons have been fully incorporated into the renormalized couplings,
the semi-classical description is valid at zero temperature.
The hot spots, which is a set of measure zero, does not affect the cyclotron period of electrons.
At non-zero temperatures,
the hot spots become hot regions with a non-zero width proportional to temperature\cite{BORGES2023169221}, and the contribution from the incoherent electrons can not be completely ignored.
Nonetheless, at low temperatures, their contribution remains sub-leading for the cyclotron period compared to the contribution away from the segments of Fermi surface outside the hot regions.
In particular, our conclusion on the non-monotonic behaviour of the cyclotron period as a function of the bare nesting angle,
which originates from outside the hot region,
remains robust as far as the size of hot regions remain much smaller than the remaining segments of Fermi surface.
However, for completeness, we consider the effect of finite temperatures below based on the scaling analysis.
We defer the more detailed study of the finite temperature correction from incoherent electrons to a future work.

In reality, the superconducting instability is inevitable in theories with non-zero bare nesting angle\cite{BORGES2023169221},
and we have to consider a non-zero temperature to be in the normal state.  
In order to understand the transport in the hot region,
one can not use the quasiparticle picture.
At temperature 
${\cal T}$ that is higher than the energy scale below which the nesting angle flows (${\cal T} > \Lambda e^{-\ell_0}$)\footnote{
In the small $v_0$ limit,
the superconducting temperature is given by
$T_c \sim
\Lambda
e^{-
\frac{a}{
\sqrt{v_0 \log 1/v_0 }}
}$
which is  is higher than the energy scale below which the nesting angle flows significantly,
$\Lambda
e^{-
\frac{b}{v_0\log(1/v_0)}}$,
where $a$ and $b$ are constants independent of $v_0$\cite{BORGES2023169221}.
}, the expression for the cyclotron period 
should be revised to
$T(v_0)=
16 \left[
T(k_i,k_c;v_0) 
+ T(k_c, k^\#;v_0)
+T(k^\#, 0;v_0)
\right]$.
Here 
$T(k_i,k_c;v_0)$  in
\eq{eq:timeitocold}
is unchanged because we are still too far away from the hot spot ($k > k_c$.)
$T(k_c, k^\#;v_0)$ 
is still given by 
\eq{eq:timecoldtostar}
except that 
$k^\# \sim {\cal T}/v$ now 
represents the momentum cut-off scale associated with temperature ${\cal T}$ (${\cal T} /4 v_0 > \Lambda e^{-\ell_0}/4 v_0 = k^*$.) 
Finally,
$ T(k^\#, 0;v_0) $
represents
the time that it takes for an incoherent electron pass through the hot region. 
In the small $v_0$ limit, the quasiparticle is only marginally destroyed 
and $ T(k^\#, 0;v_0)$ can be estimated to be $ T(k^\#, 0;v_0)  \sim  \frac{ k^\# }{ e B_0  \left( k^\#/k_c \right)^{\alpha_1} }$.
Here, the Fermi velocity has the form $(k^\#/k_c)^{\alpha_1}$ because the dynamics of the hot region at temperature ${\cal T}$ now replaces the zero-temperarture dynamics that previously contained the nesting angle profile ($k^\# > k^*$.)

In the zero-temperature superconducting state\cite{BORGES2023169221},
one has to include the  effect of the pair condensate to describe the dynamics of the Bogoliubov quasiparticles.
Suppose that the ground state has the d-wave pairing\cite{SCALAPINO}
with a momentum-dependent pairing wavefunction $\Delta_k$.
For the physical case with $N_c=2$ and $N_f = 1$,
we need to add the following action 
for quasiparticles 
in segments $1$ and $5$,
\begin{equation}
    S'^{1,5} = \sum_{\sigma,\sigma'=\uparrow,\downarrow}\int\frac{\dd \omega \dd^2 k}{(2\pi)^3}\left\lbrace\psi^*_{1,\sigma}(\omega,\vec{k})\Delta^{\sigma\sigma'}_k\psi^*_{5,\sigma'}(-\omega,-\vec{k}) + \psi_{5,\sigma}(-\omega,-\vec{k})\left.\Delta^{\dagger}_k\right.^{\sigma\sigma'}\psi_{1,\sigma'}(\omega,\vec{k})\right\rbrace,
    \label{eq:pairingwvfterm}
\end{equation}
where
$\Delta^{\sigma\sigma'}_k = \Delta_k \epsilon^{\sigma\sigma'}$ 
    with
    $\epsilon = \left(
    \begin{matrix}
    0 & 1 \\
    -1 & 0
    \end{matrix}
    \right)$.
The derivation of this equation can be found in Appendix \ref{app:HubbStrat}. Eqs. \eqref{eq:quadraticaction} and \eqref{eq:pairingwvfterm}
can be combined into an action
of a spinor field that represents
Bogoliubov quasiparticles in the superconducting state,
\begin{equation}
    S^{1,5}_{\mathrm{kin}}+S'^{1,5} = \int\frac{\dd \omega \dd^2 k}{(2\pi)^3} \Bar{\Uppsi}(\omega,\vec{k})\left\lbrace i \omega \Gamma^0 +i\mathcal{V}_{F,k}(v_k k_x + k_y)\Gamma^1 - i\Delta_k \Gamma^2\right\rbrace\Uppsi(\omega,\vec{k}),
    \label{eq:epsilonpairing}
\end{equation}
where $S^{1,5}_{\mathrm{kin}}$ are the terms of the kinematic action corresponding to electrons in hot spots 1 and 5, and
$\Uppsi^T(\omega,\vec{k}) = \left(\psi_{1,\uparrow}(\omega,\vec{k}),\psi_{1,\downarrow}(\omega,\vec{k}),\psi^*_{5,\downarrow}(-\omega,-\vec{k}),-\psi^*_{5,\uparrow}(-\omega,-\vec{k})  \right)$
is a $4$-component spinor.
$\Bar{\Uppsi}(\omega,\vec{k}) = \Uppsi^\dagger(\omega,\vec{k})\Gamma^0$
with
$ \Gamma^0 = \sigma_y\otimes\mathds{1}_2$, 
$\Gamma^1 = \sigma_x\otimes\mathds{1}_2$, 
$\Gamma^2 = \sigma_z\otimes\mathds{1}_2$
being $4 \times 4$ gamma matrices,
where the first Pauli matrices act on the Nambu spinor basis
and the second Pauli matrices act on the spin space.
Using the same transform as in Eq. \eqref{eq:spinordef} for the Dirac spinor $\Uppsi$ in the hybrid momentum-spacetime $(t,r,k)$,
Eq. \eqref{eq:epsilonpairing} can be written as
\begin{equation}
S^{1,5}_{\mathrm{kin}}+S'^{1,5} = \int\frac{\dd k}{2\pi}\int \dd t \dd r \abs{\mathfrak{e}}\Bar{\Uppsi}(t,r,k)\left\lbrace\Gamma^0 \mathfrak{e}_{0}^{\ t}D_t +  \Gamma^1\mathfrak{e}_{1}^{\ r}D_r\right\rbrace\Uppsi(t,r,k).
\label{eq:covariantaction}
\end{equation}
Here, the vielbein and the U(1) gauge field are unchanged, but the pairing term gives rise to a complex spin connection,
$ \omega_{t,02} = 4i\Delta_k$.
It will be of interest to find geometric interpretation of physical observables in the superconducting state.
\section{Conclusion and Outlook}
\label{sec:con}

In this paper, 
we show that
the momentum-dependent quantum correction that  dilates frequency of electron anisotropically on the Fermi surface gives rise to 
a curved momentum-spacetime
for low-energy quasiparticles
in the 2+1 dimensional antiferromagnetic  quantum critical metal.
The non-trivial dependence of the emergent geometry on the  shape of the Fermi surface causes a non-monotonic dependence of the cyclotron frequency on the bare nesting angle of the Fermi surface.
With increasing nesting angle, 
the stronger disparity in the strength of quantum correction in different parts of Fermi surface makes the momentum-dependent red shift more singular at the hot spots.
This creates a possibility of realizing an analogous black hole horizon at the hot spots where the motion of quasiparticle tend to freeze beyond a critical bare nesting angle of the Fermi surface.
However, this analogous horizon does not lead to a vanishing cyclotron frequency 
because the metric 
 in the vicinity of the hot spots is modified 
by thermal effects 
above the superconducting transition temperature.
Our prediction can be in principle tested through 
 a measurement of the cyclotron frequency as a function of the nesting angle near the hot spots in one of those materials.

We close with some discussions and outlook:

\begin{itemize}
\item 
We note that it is not necessary to use the perspective of curved momentum-spacetime to understand how the cyclotron period depends on the bare nesting angle.
One can attribute the momentum-dependent quantum effect either to the Fermi velocity or to the background geometry.
In the non-geometric picture, one can  explain the same physics by introducing a momentum-dependent Fermi velocity in a flat momentum-spacetime.
However, the main advantage of the geometric perspective is a conceptual one.
Even in the non-geometric picture, the origin of the momentum-dependent Fermi velocity is the red shift that dilates the frequency in a momentum-dependent manner.
What makes electrons slower near the hot spots is not a modification of the dispersion but the slowing down of their clocks from the perspective of electrons in the cold region.
This makes it more natural to understand the dynamics of quasiparticles from the geometric perspective.
    \item 
    Although we considered the example of antiferromagnetic quantum critical metal,
    the emergence of a curved momentum space is a general phenomenon in metals where quantum corrections depend strongly on momentum along the Fermi surface.
    In any quantum critical metal associated with an order parameter carrying a non-zero momentum such as charge density wave criticality, we expect that a similar curved momentum-spacetime will emerge near the hot spots.
    Conversely, there are quantum critical metals with cold spots at which there is an enhanced blue shift.
    In the Ising-nematic quantum critical metal, quantum corrections vanish at discrete points on the Fermi surface due to the form factor of the order parameter that changes sign around the Fermi surface.
    It would be interesting to understand the dynamics of electrons near the cold spots in terms of the emergent momentum-space geometry.
%
%
    \item 
   In this paper, we mainly  considered the zero temperature limit in which the quasiparticle description is valid almost everywhere on the Fermi surface.
   At finite temperatures, the contribution of the incoherent electrons in the vicinity of the hot spots is not negligible. 
   It is of great interest to understand the dynamics of incoherent electrons in a more general setting.
   In particular, it is curious how  a momentum-space geometry defined locally in the momentum space interplay with large angle scatterings that are non-local in the momentum space.
  %

    \item 
    In the present study, the momentum dependent red-shift is the primary mechanism by which the Fermi velocity acquires strong momentum dependence near the hot spots.
    In general, other effects, such as the renormalization of the fermion dispersion, can further renormalize the dynamics of electrons. 
    In the antiferromagnetic quantum critical metal, its effect is sub-leading in the limit that the nesting angle is small.
    It would be of interest to find examples where a more general form of curved momentum-spacetime emerge from such quantum corrections.

\end{itemize}

\newpage

\appendix
\section{
The origin of the momentum-dependent coupling functions
}
\label{app:FRG}

In this appendix,
we review the functional renormalization group flow 
that gives rise to the momentum-dependence for the coupling
functions\cite{BORGES2023169221}.
%
%
%
%
The power of the floating energy scale $\mu$ in front of each coupling function in Eq. \eqref{eq:LukewarmAction}
denotes its scaling dimension 
under the interaction driven scaling\cite{SHOUVIK2} 
that is exact in the limit that the nesting angle is small\cite{SCHLIEF}.
The four-fermion coupling function $\lambda$, 
which has scaling dimension $-1$, should be included in the renormalizable theory 
because it is promoted to a marginal coupling with the help of the Fermi momentum\cite{BORGES2023169221}.
On the contrary, 
quartic boson couplings
are strictly irrelevant under the interaction-driven scaling
and can be dropped in the low-energy theory\cite{SCHLIEF}.
The low-energy theory does not include the 
 quadratic action of the boson either 
 because the local kinetic term is irrelevant under the interaction driven scaling\footnote{
Physically, this implies that the two-point function of the boson is entirely determined by
quantum corrections at low energies.}.
The absence of the bare boson kinetic term in the action gives us the freedom to rescale the boson field  to tune the magnitude of the Yukawa coupling at the hot spots.
The physics does not depend on this choice, but
it is convenient to choose the normalization 
so that $g_{0,0} = \sqrt{\pi v_0/2}$.
With this choice, the boson self-energy is `canonically' normalized
to 
\begin{equation}
D({\bf q})=\frac{1}{
|q_0| + c 
\left( |q_x| + | q_y| \right)},
\end{equation}
where
\begin{equation}
c = \sqrt{\frac{v_0}{8 N_c N_f}\log\left(\frac{1}{v_0}\right)}
\end{equation}
is the speed of the dressed collective mode 
that is entirely determined from the nesting angle at the hot spots\cite{SCHLIEF}.
We also have the freedom to choose the scale of frequency relative to momentum.
We first follow Ref. \cite{BORGES2023169221} to choose the unit of frequency such that $V_{\mathrm{F},0} =1$ at all scales\footnote{
With this choice, $V_{F,k}$ at $k \neq 0$ represents
the Fermi velocity measured in the unit of the Fermi velocity at the hot spots.}.
Since the electrons at the hot spots receive quantum corrections down to the zero energy limit, 
keeping $V_{F,k}=1$ 
in  \eq{eq:VFk}
at all $\mu$
requires a continuous redefinition of $k_0$ relative to the bare frequency
as the energy scale is lowered.
Since we are using one global clock according to which the electrons at the hot spots have a fixed velocity, cold electrons away from the hot spots appear to be moving faster with this choice of frequency unit.

The evolution of  the momentum-dependent vertex function (See Eq. \eqref{eq:VFk}) with decreasing $\mu$ defines the functional renormalization group flow of the coupling functions\cite{BORGES2023169221}.
In the space of coupling functions,
an {\it interacting} fixed point arises at
\begin{equation}
    v_k = 0, \quad V_{F,k} = 1, \quad g_{k',k} = 0, \quad \lambda^{\spmqty{N_1 & N_ 2 \\ N_4 & N_3}; \spmqty{\sigma_1 & \sigma_2 \\ \sigma_4 & \sigma_3}}_{\spmqty{k_1 & k_2\\ k_1 + k_2 - k_3 & k_3}} = 0
    \label{eq:fxpt}
\end{equation}
with $g^2_{k',k}/v_k = \pi/2$. 
Although the couplings vanish at the fixed point, quantum corrections remain non-trivial due to the vanishing nesting angle 
at the fixed point.
The anomalous dimension of the boson is controlled by 
$g^2_{k',k}/v_k \sim O(1)$
and \eq{eq:fxpt} is far from the Gaussian fixed point. 
For theories with non-zero nesting angles at UV, 
the coupling functions undergo a non-trivial renormalization group flow as energy is lowered.
At low energies,
the coupling functions acquire momentum dependence under the functional renormalization group flow 
because quantum corrections depend on momentum along the Fermi surface.
In this paper, we consider the case,
where $v_k$ and $V_{F,k}$ are independent of momentum  at UV cutoff scale $\Lambda$
: $v_k(\ell=0)=v_0(0)$
and $V_{F,k}(\ell=0)=1$,
where $\ell \equiv \log \Lambda/\mu$ is the logarithmic length scale.
For different UV theories, the exact profiles of the renormalized coupling functions are different, but the part that is singular at the hot spots is universal\cite{BORGES2023169221}.

\begin{figure}[t]
	\centering
	\begin{subfigure}[b]{0.49\linewidth}
		\centering
		\includegraphics[scale=1]{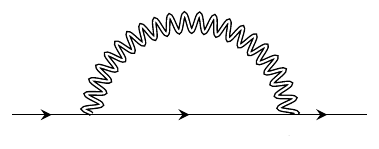}
        \includegraphics[scale=1]{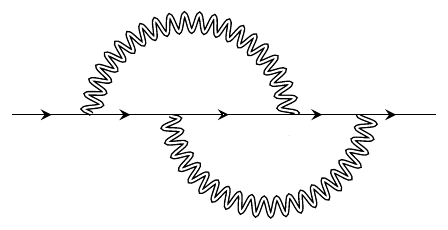}\vspace{1.5cm}
		\caption{\label{fig:1lfse}}
	\end{subfigure}
	\begin{subfigure}[b]{0.49\linewidth}
		\centering
		\includegraphics[scale=0.8]{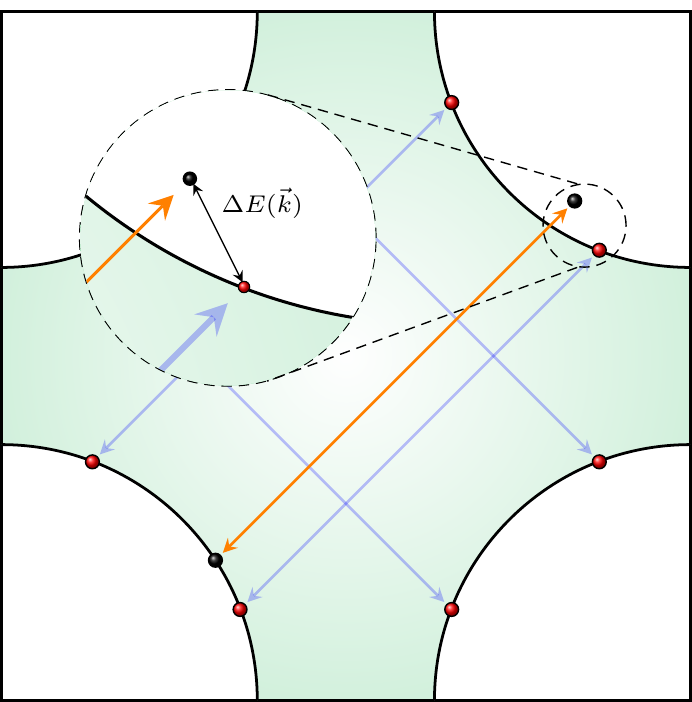}
		\caption{\label{fig:crossoverscale}}
	\end{subfigure}
	\caption{
	({\color{blue}$a$}) 
The leading quantum corrections that dress the quadratic action of electron in the limit that the nesting angle is small.
The double wiggly line represents the fully dressed boson propagator obtained from the Schwinger-Dyson equation\cite{SCHLIEF}.
({\color{blue}$b$}) 
The virtual electron created when an electron near hot spot $5$ emits a boson with zero energy.
If the electron is right at the hot spot $5$, it is scattered onto the hot spot $8$ on the Fermi surface.
If the electron is away from the hot spot,
it is scattered into a state away from the Fermi surface.
The non-zero energy 
of the virtual particle 
(schematically called $\Delta E(\vec k)$ here)
cuts off the IR singularity of the self-energy. The form of $\Delta E(\vec{k})$ here depends on the channel of interaction. For example, for the one-loop fermion self-energy in (${\color{blue} a}$) $\Delta E (\vec{k}) = E^{(1L)}$ and for the two-loop fermion self energy, $\Delta E (\vec{k}) = E^{(2L)}$ (See Eq. \eqref{eq:crossovermscales}.)
}
\end{figure}
\begin{figure}[t]
    \centering
    \includegraphics[scale=1]{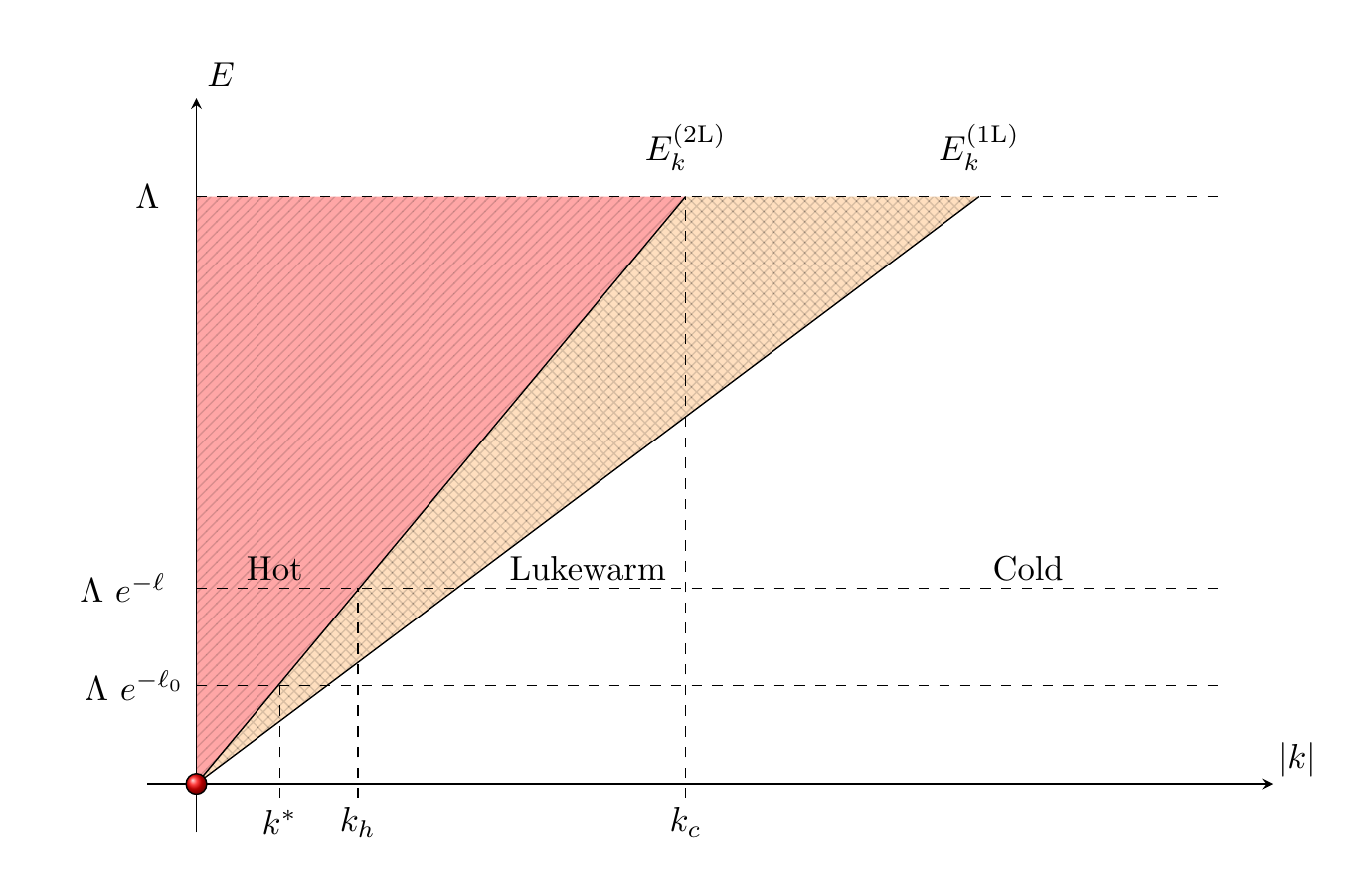}
    \caption{
    Crossover energy scales associated with the two-loop fermion self-energy 
    ($E^{(2L)}_k$) and the one-loop fermion self-energy ($E^{(1L)}_k$), respectively. 
    At a small nesting angle, 
    the flow of couplings between 
    the two crossover energy scales is negligible,
    and one can use the approximation 
    in which both quantum corrections turn off 
    simultaneously below $E^{(2L)}_k$. 
    At energy scales $\mu=\Lambda e^{-\ell}$,
    the momentum space is divided into three regions depending on the magnitude of $E^{(2L)}_k$ 
    relative to $\mu$ and $\Lambda$.
    $\mu_0 \equiv \Lambda e^{-\ell_0}$ denotes the energy scale below which the flow of the nesting angle becomes important.
    In $k< k^*$,
    the shape of the Fermi surface is significantly renormalized.
}
    \label{fig:IRscales}
\end{figure}

In the small nesting angle limit, only the one-loop and two-loop fermion self-energy corrections are important
 for the quadratic action of electrons (Fig. \ref{fig:1lfse}). 
If the external electron is right 
at a hot spot,
the electron can be scattered to another hot spot by emitting virtual bosons with zero energy. 
Since all virtual particles can have zero energy in the loop, a logarithmic singularity arises for electrons at hot spots. 
If the external electron is away from hot spots, 
virtual particles are forced to carry a non-zero energy.
The energy, 
which is proportional to the deviation of momentum away from the hot spot and the nesting angle,
cuts off the infrared singularity 
as is illustrated in		 \fig{fig:crossoverscale}.
Because of this, electrons away from hot spots become decoupled from spin fluctuations at sufficiently low energies.
The crossover energy scales 
 for the one-loop and two-loop fermion self-energy 
of the electron on the Fermi surface
with momentum 
$k=k_N$ relative to the hot spots
are given by
\begin{equation}
   E^{(1L)} = 2 c v_k\abs{k}, \quad E^{(2L)} = 4 V_{F,k} v_k \abs{k}.
    \label{eq:crossovermscales}
\end{equation}
These crossover energy scales as functions of $k$ are shown in     \fig{fig:IRscales}.
Each quantum correction turns off at energies  far below $E^{(1L)}_k$ and $E^{(2L)}_k$, respectively.  
The logarithmic length scales associated with Eq. \eqref{eq:crossovermscales} as \begin{equation}
    \ell^{(1L)}_{k} = \log\left(\frac{\Lambda}
    {
2 c v_k |k|
    }\right), \quad \ell^{(2L)}_{k} = \log\left(\frac{\Lambda}{
4 V_{F,k} v_k |k|
    }\right).
    \label{eq:lengthmscales}
\end{equation}
In the small $v$ limit,
the flow of couplings is negligible in 
$\ell^{(2L)}_{k}  < \ell < \ell^{(1L)}_{k} $. 
Therefore, 
to the leading order in $v$,
one can merge the two crossovers into one :
for $\ell \gg  \ell^{(2L)}_{k}$,
both the one-loop and two-loop quantum corrections renormalize the Fermi velocity and the nesting angle,
and 
for $\ell \ll  \ell^{(2L)}_{k}$,
both quantum corrections turn off.
Because the quantum corrections that renormalize the coupling functions turn off at different energy scales at different momenta, 
the renormalized coupling functions end up acquiring non-trivial momentum dependences at low energies as is shown in Eqs.
   \ref{eq:nestinAng} and \ref{eq:FermiVel}.
   Electrons closer to the hot spots remain coupled with spin fluctuations for a larger window of energy scales,
   and their Fermi velocity becomes smaller compared to electrons far away from hot spots.
   For the same reason, the electrons near the hot spots exhibit the stronger emergent nesting, that is the smaller renormalized nesting angle.

\section{Spinors in curved spacetime}
\label{app:spinors}

In this appendix, 
we review the background material for the theory of spinor in curved spacetimes.
For concreteness, 
we consider 
four-dimensional spacetimes, but
the discussion can be generalized to any dimension
(for more details, see Ref. \cite{birrell_davies_1982} for example). 
Suppose manifold $M$ is endowed with metric $g_{\mu\nu}$. 
One can define 
a set of orthonormal basis 
that spans the space of one-forms through which
the metric two-form can be written as
%
 \begin{equation}
    g_{\mu\nu}\dd x^\mu \otimes\dd x^\nu = \eta_{ab}\hat{\theta}^{a} \otimes \hat{\theta}^{b},
\end{equation}
where
$\hat{\theta}^a = \mathfrak{e}^{a}_{\ \mu}\dd x^\mu$
is the orthonormal basis,
$\mathfrak{e}^{a}_{\ \mu}$ is the vielbein
and
%
%
$\eta_{ab} = \text{diag}(-1,+1,+1,+1)$\footnote{
For Euclidean signature, the Minkowskian metric $\eta_{ab}$ is replaced with $\delta_{ab}=\text{diag}(+1,+1,+1,+1)$.
In the following two paragraphs, we use the Einstein convention for the Greek and Latin indices. 
}. 
One can also introduce a set of vectors that is dual to
$\{\hat{\theta}^{a}\}$ 
through the relation
$\langle\hat{\theta}^a, \hat{e}_b\rangle = \delta^{a}_{\ b}$,
where
$\hat{e}_a = \mathfrak{e}_{a}^{\ \mu}\partial_\mu$.
%
%
The bases $\{\hat{e}_a\}$ and $\{\hat{\theta}^a\}$ are called the {\it non-coordinate bases}. 
There is clearly the freedom to rotate the orthonormal basis through the local Lorentz transformations. 
A relativistic fermion forms a spinor representation under this local Lorentz transformation.
From the requirement that the action should be invariant under the local Lorentz transformation, 
the action for a Dirac spinor with mass $m$ coupled to a gauge field $\mathcal{A}_\mu(x)$ is
written as
\begin{equation}
    \Gamma_\psi = \int_M \sqrt{\abs{g}} \dd^4 x ~ \Bar{\psi}(x)\gamma^c\mathfrak{e}_{c}^{\ \mu}\left(\partial_\mu + \mathcal{A}_\mu(x) + \frac{i}{2}\omega_{\mu,ab}\Sigma^{ab} + m\right)\psi(x).
\end{equation}
Here, 
$\sqrt{\abs{g}} = \abs{\mathfrak{e}}$ with  $\mathfrak{e} = \text{det }\mathfrak{e}^a_{\ \mu}$.  
$\bar{\psi} = \psi^\dagger \gamma^0$, where the gamma matrices $\gamma^a$ satisfy the Clifford algebra $\{\gamma^a , \gamma^b\} = 2\eta^{ab}$ that furnishes the 
 spinor representation of local Lorentz transformation.
$\Sigma^{ab}=\frac{i}{4}\left[\gamma^a,\gamma^b\right]$ is the generator of the local Lorentz transformations. 
The matrix-valued one form with elements $\omega_{\mu, ab} = (\omega_{ab})_\mu$ is the spin connection that acts as the gauge connection for the local Lorentz transformation.
The connection one-form  
and the vielbein completely determines the torsion and curvature of the spacetime through
the {\it Cartan structure equations},
\begin{subequations}\label{eq:Cartan}
    \begin{align}
        \dd \hat{\theta}^a + \omega^{a}_{\ b}\wedge\hat{\theta}^{b} = & T^{a},
        \label{eq:Torsion} \\
        \dd \omega^{a}_{\ b} + \omega^{a}_{\ c}\wedge\omega^{c}_{\ b} = & R^{a}_{\ b}, \label{eq:Curvature}
    \end{align}
where $\dd$ is the exterior derivative, $\wedge$ is the wedge product,
$T^{a}$ is the torsion two-form and 
$R^{a}_{\ b}$ is the curvature two-form. 
$T^a$ and $R^{a}_{\ b}$ are gauge covariant measures that characterize the geometry of the manifold $M$.
The torsion measures the screw or twist on a frame when parallel transported along two directions in the manifold $M$. 
The curvature is a measure of the holonomy (gauge flux) acquired by a vector transported parallely around a loop.

\end{subequations}
\section{Schwarzschild Geodesics}
\label{app:SchwarzschildBH}

In this appendix,
we briefly review the Schwarzschild spacetime
that describes a black hole
and the motion of a free falling object
(for a more complete review, see \cite{wald2010general} for example).
The Schwarzschild metric is an exact solution to Einstein field equations, which describes the spacetime outside  
a static, 
neutral
and
spherically symmetric compact object.
In the spherical coordinate, 
the line element for the Schwarzschild metric can be written as
\begin{equation}
 ds^2= \left(1-\frac{r_s}{r}\right)c^2\dd t^2 - \frac{\dd r^2}{1-\frac{r_s}{r}}-r^2\dd \theta^2 - r^2 \sin^2\theta \dd \phi^2,
    \label{eq:Schwarzschild}
\end{equation}
where $M$ is the mass
and 
$c$ is the sped of light. $t$ is the time coordinate 
 that describes the proper time of an asymptotic observer who is at 
 $r=\infty$, where
$r$ is the radial coordinate.
\begin{equation}
    r_s = \frac{2 G M}{c^2}
\end{equation}
is the Schwarzschild radius, where $G$ is the gravitational constant. 
When the radius of the compact object is smaller than $r_s$, 
it describes a black hole 
with an event horizon at 
 $r=r_s$ inside of which the future light cone always points toward $r=0$, and thus no escape.
As the free falling object approaches the horizon from the outside, 
the lapse of its proper time is infinitely slowed down as compared to that of the proper time of an asymptotic observer.
This gives rise to a critical slowdown of the motion of the free falling object as observed by the asymptotic observer.
For an object that 
 falls radially to the black hole at a fixed angle 
(say $\theta = \pi/2$ and $\phi=0$),
the geodesic equation of motion gives
%
\begin{equation}
    t \sim
   \frac{r_s}{c}
   \log\left(\frac{\abs{\sqrt{\frac{r}{r_s}}+1}}{\sqrt{\frac{r}{r_s}}-1}\right)
    \label{eq:geodesic}
\end{equation}
near the horizon.
As $r$ approaches $r_s$,
$t$ diverges logarithmically.
This implies that the asymptotic observer will never see the passing of the free falling object across the horizon.
%
%
\section{
The pairing term for Bogoliubov quasiparticles}
\label{app:HubbStrat}

In this appendix we derive Eq. \eqref{eq:pairingwvfterm}.
We begin by writing the pairing interaction between
electrons in segments $1$ and $5$
(See Fig. \ref{fig:FermiSurface}),
\begin{equation}
\begin{aligned}
    S_{4\mathrm{f}}^{\spmqty{1 & 5 \\ 1 & 5}} = \frac{1}{4\mu}\sum_{\{\sigma_i\}}\int\frac{\dd\omega_p\dd\omega_k\dd\omega_q~\dd^2p~\dd^2k~\dd^2q}{(2\pi)^9}& \psi^*_{1,\sigma_1}\left(\omega_p + \frac{\omega_q}{2},\vec{p} + \frac{\vec{q}}{2}\right)\psi^*_{5,\sigma_2}\left(-\omega_p + \frac{\omega_q}{2},-\vec{p} + \frac{\vec{q}}{2}\right)\times
    \\ & \lambda_{\spmqty{p + \frac{q}{2}& - p + \frac{q}{2}\\ k + \frac{q}{2} & - k + \frac{q}{2}}}^{\spmqty{1 & 5 \\ 1 & 5};\spmqty{\sigma_1 & \sigma_2 \\ \sigma_4 & \sigma_3}}\psi_{5,\sigma_3}\left(-\omega_k + \frac{\omega_q}{2},-\vec{k} + \frac{\vec{q}}{2}\right)\psi_{1,\sigma_4}\left(\omega_k + \frac{\omega_q}{2},\vec{k} + \frac{\vec{q}}{2}\right).
\end{aligned}
\label{eq:15fouraction}
\end{equation}
Here, $\lambda$ is the four fermion coupling function generated from the critical spin fluctuations.
The specific momentum-dependence of the coupling function is not important for us.
The functional renormalization group analysis shows that the strongest attractive interaction is generated in the spin-singlet and d-wave channel,
and the significant pairing interaction is generated not only for the electrons at the hot spots but also for electrons that are far away from the hot spots
\cite{BORGES2023169221}.
We are using the frequency defined in Eq. \eqref{eq:frequencyrescaling}.
Performing a Hubbard-Stratonovich transformation
on the quartic interaction yields
\begin{equation}
\begin{aligned}
    & e^{
    -S_{4\mathrm{f}}^{\spmqty{1 & 5 \\ 1 & 5}}} = \int\mathcal{D}\Delta\mathcal{D}\Delta^*\exp{\Bigg\lbrace}
    - \sum_{\{\sigma_i\}}\int\left(\prod_{l=p,q,k,q'}\frac{\dd\omega_l~\dd^2 l}{(2\pi)^{3}}\right){\Bigg[}\left.\tilde{\Delta}^\dagger\right.^{\sigma_1\sigma_2}_{(\omega_p,\vec{p};\omega_q,\vec{q})}\left[-\bar{\lambda}\right]^{\sigma_1\sigma_2;\sigma_3\sigma_4}_{(\omega_p,\vec{p},\omega_q,\vec{q};\omega_k,\vec{k},\omega_{q'},\vec{q'})}\tilde{\Delta}^{\sigma_3\sigma_4}_{(\omega_k,\vec{k};\omega_{q'},\vec{q'})}
    \\ & -\psi^*_{1,\sigma_1}\left(\omega_p + \frac{\omega_q}{2},\vec{p} + \frac{\vec{q}}{2}\right)\psi^*_{5,\sigma_2}\left(-\omega_p + \frac{\omega_q}{2},-\vec{p} + \frac{\vec{q}}{2}\right)\left[-\bar{\lambda}\right]^{\sigma_1\sigma_2;\sigma_3\sigma_4}_{(\omega_p,\vec{p},\omega_q,\vec{q};\omega_k,\vec{k},\omega_{q'},\vec{q'})}\tilde{\Delta}^{\sigma_3\sigma_4}_{(\omega_k,\vec{k};\omega_{q'},\vec{q'})}
    \\ & -\left.\tilde{\Delta}^\dagger\right.^{\sigma_1\sigma_2}_{(\omega_p,\vec{p};\omega_q,\vec{q})}\left[-\bar{\lambda}\right]^{\sigma_1\sigma_2;\sigma_3\sigma_4}_{(\omega_p,\vec{p},\omega_q,\vec{q};\omega_k,\vec{k},\omega_{q'},\vec{q'})}\psi_{5,\sigma_3}\left(-\omega_k + \frac{\omega_{q'}}{2},-\vec{k} + \frac{\vec{q'}}{2}\right)\psi_{1,\sigma_4}\left(\omega_k + \frac{\omega_{q'}}{2},\vec{k} + \frac{\vec{q'}}{2}\right){\Bigg]}{\Bigg\rbrace},
\end{aligned}
\label{eq:HubbStr}
\end{equation}
where 
we use the matrix notation $\left[\bar{\lambda}\right]^{\sigma_1\sigma_2;\sigma_3\sigma_4}_{(\omega_p,\vec{p},\omega_q,\vec{q};\omega_k,\vec{k},\omega_{q'},\vec{q'})} = \frac{(2\pi)^3\delta(\omega_p-\omega_{q'})\delta^{(2)}(\vec{q}-\vec{q'})}{4\mu}\lambda_{\spmqty{p + \frac{q}{2} & - p + \frac{q}{2}\\ k + \frac{q'}{2} & - k + \frac{q'}{2}}}^{\spmqty{1 & 5 \\ 1 & 5};\spmqty{\sigma_1 & \sigma_2 \\ \sigma_4 & \sigma_3}}$
\footnote{The matrix elements of $\bar{\lambda}$ can be written as $\bar{\lambda}_{mn}$, where $m \equiv (\sigma_1\sigma_2;\omega_p,\vec{p},\omega_q,\vec{q})$ and $n \equiv (\sigma_3\sigma_4;\omega_k,\vec{k},\omega_q',\vec{q'})$.}. 
%
%
%
%
%
Writing the static component of the order parameter with zero center of mass momentum 
as
$\tilde{\Delta}^{\sigma_1\sigma_2}_{(\omega_p,\vec{p};\omega_q,\vec{q})} = 
(2\pi)^3\delta(\omega_q)\delta^{(2)}(\vec{q}) \tilde{\Delta}^{\sigma_1\sigma_2}_{(\omega_p,\vec{p})}$
and making a further transformation through
%
%
%
%
\begin{equation}
\Delta^{\sigma_1\sigma_2}_{(\omega_p,\vec{p})} = \sum_{\sigma_3\sigma_4}\int\frac{\dd\omega_k\dd^2k}{(2\pi)^3}\frac{\lambda_{\spmqty{p & - p \\ k & - k }}^{\spmqty{1 & 5 \\ 1 & 5};\spmqty{\sigma_1 & \sigma_2 \\ \sigma_4 & \sigma_3}}}{4\mu}\tilde{\Delta}^{\sigma_3\sigma_4}_{(\omega_k,\vec{k})},
    \label{eq:paringampdef}
\end{equation}
we obtain the pairing term for the Bogoliubov quasiparticles,
\begin{equation}
\begin{aligned}
    & e^{
    -S_{4\mathrm{f}}^{\spmqty{1 & 5 \\ 1 & 5}}} = \int\mathcal{D}\Delta\mathcal{D}\Delta^*\exp{\Bigg\lbrace}
    - \sum_{\{\sigma_i\}}\int\frac{\dd\omega_p~\dd^2 p\dd\omega_k~\dd^2 k}{(2\pi)^{6}}\left.\Delta^\dagger\right.^{\sigma_1\sigma_2}_{(\omega_p,\vec{p})}[\mathfrak{G}^{-1}]^{\spmqty{\sigma_1 & \sigma_2 \\ \sigma_4 & \sigma_3}}_{\spmqty{p & - p \\ k & -k}}\Delta^{\sigma_3\sigma_4}_{(\omega_k,\vec{k})}
    \\ & - \sum_{\sigma_1,\sigma_2}\int\frac{\dd\omega_p~\dd^2 p}{(2\pi)^{3}}\left[\psi^*_{1,\sigma_1}\left(\omega_p,\vec{p}\right)\Delta^{\sigma_1\sigma_2}_{(\omega_p,\vec{p})}\psi^*_{5,\sigma_2}\left(-\omega_p,-\vec{p}\right) + \psi_{5,\sigma_1}\left(-\omega_p,-\vec{p}\right)\left.\Delta^\dagger\right.^{\sigma_1\sigma_2}_{(\omega_p,\vec{p})}\psi_{1,\sigma_2}\left(\omega_p,\vec{p}\right)\right]{\Bigg\rbrace},
\end{aligned}
\label{eq:HubbStrFinal}
\end{equation}
where $[\mathfrak{G}^{-1}]^{\spmqty{\sigma_1 & \sigma_2 \\ \sigma_4 & \sigma_3}}_{\spmqty{p & - p \\ k & -k}} = 4\mu(2\pi)^3\delta(0)\delta^{(2)}(\vec{0})\left[-\lambda^{-1}\right]^{\spmqty{1 & 5 \\ 1 & 5};\spmqty{\sigma_1 & \sigma_2 \\ \sigma_4 & \sigma_3}}_{\spmqty{p & - p \\ k & -k}}$.
Once the momentum-dependent pairing amplitude is determined from the saddle-point equation,
it together with the vielbein sets the  background spacetime 
on which the Bogoliubov quasiparticles propagate.


\section*{Acknowledgement}

This research was supported by the Natural Sciences 
and Engineering Research Council of
Canada. Research at the Perimeter Institute is supported in part by the
Government of Canada through Industry Canada, and by the Province of
Ontario through the Ministry of Research and Information.

\bibliographystyle{apsrev4-2}

\bibliography{references}

\end{document}